\documentclass[useAMS,usenatbib,referee,12pt]{article}
\usepackage{graphicx}
\usepackage{enumerate}
\usepackage{url} % not crucial - just used below for the URL 
\usepackage[strings]{underscore}
\usepackage{natbib}
\usepackage{mathrsfs}
\usepackage{bm}
\usepackage{ dsfont }
\usepackage{amsmath, amssymb}
\usepackage{setspace}
\usepackage{xcolor}
\usepackage{rotating}
\usepackage{mleftright}
\usepackage{xparse}
\usepackage{booktabs}
\usepackage{multirow}
\usepackage{adjustbox}
\usepackage{longtable}
\usepackage[tableposition=below]{caption}
\definecolor{light-gray}{gray}{0.95} 
\definecolor{mypink}{RGB}{255, 0, 255}
\DeclareMathOperator{\EX}{\mathbb{E}}% expected value
\NewDocumentCommand{\evalat}{sO{\big}mm}{%
  \IfBooleanTF{#1}
   {\mleft. #3 \mright|_{#4}}
   {#3#2|_{#4}}%
}

\newcommand{\blind}{1}

\begin{document}

\bibliographystyle{apalike}

\def\spacingset#1{\renewcommand{\baselinestretch}%
{#1}\small\normalsize} \spacingset{1}

%%%%%%%%%%%%%%%%%%%%%%%%%%%%%%%%%%%%%%%%%%%%%%%%%%%%%%%%%%%%%%%%%%%%%%%%%%%%%%

\if1\blind
{
  \title{\bf{A New Class of Mark-Specific Proportional Hazards Models for Recurrent Events: Application to Opioid Refills Among Post-Surgical Patients}}
  \author{Eileen Yang\footnote{Email: ejyang@umich.edu}\\
    Department of Biostatistics, University of Michigan, Ann Arbor, MI, 48109\\
    and \\
    Donglin Zeng\footnote{Email: dzeng@umich.edu} \\
    Department of Biostatistics, University of Michigan, Ann Arbor, MI, 48109\\
    and \\
    Mark Bicket\footnote{Email: mbicket@med.umich.edu}\\
    Department of Anesthesiology, University of Michigan, Ann Arbor, MI, 48109\\
    and \\
    Yi Li\footnote{Corresponding author. Email: yili@umich.edu} \\
    Department of Biostatistics, University of Michigan, Ann Arbor, MI, 48109\\
    }
  \maketitle
} \fi

\if0\blind
{
  \bigskip
  \bigskip
  \bigskip
  \begin{center}
    {\Large \bf{A New Class of Mark-Specific Proportional Hazards Models for Recurrent Events: Application to Opioid Refills Among Post-Surgical Patients}}
\end{center}
  \medskip
} \fi

\bigskip
\begin{abstract}

Prescription opioids relieve moderate-to-severe pain after surgery, but overprescription can lead to misuse and overdose. Understanding factors associated with post-surgical opioid refills is crucial for improving pain management and reducing opioid-related harms. Conventional methods often fail to account for refill size or dosage and capture patient risk dynamics. We address this gap by treating dosage as a continuously varying mark for each refill event and proposing a new class of mark-specific proportional hazards models for recurrent events. Our marginal model, developed on the gap-time scale with a dual weighting scheme, accommodates event proximity to dosage of interest while accounting for the informative number of recurrences. We establish consistency and asymptotic normality of the estimator and provide a sandwich variance estimator for robust inference. Simulations show improved finite-sample performance over competing methods. We apply the model to data from 
the Michigan Surgical Quality Collaborative and Michigan Automated Prescription System. 
Results show that high BMI, smoking, cancer, and open surgery increase hazards of high-dosage refills, while inpatient surgeries elevate refill hazards across all dosages. Black race is associated with higher hazards of low-dosage but lower hazards of high-dosage refills. These findings may inform personalized, dosage-specific pain management strategies.
 
\end{abstract}
\noindent%
\vfill

\newpage
\spacingset{1.9} % DON'T change the spacing!

\section{Introduction}
\label{sec:intro}

% Para 1
Opioid misuse, which has claimed more than 645,000 lives in the U.S. over the past two decades, often originates from exposure to opioids prescribed by clinicians to treat pain \citep{cdc_understanding_2024, biancuzzi_opioid_2022}. 
One of the most common settings in which opioids are prescribed is postoperative pain management \citep{hinther_chronic_2019}. As proper pain management is crucial for postoperative healing, recovery, and successful surgical outcomes, opioids remain a common choice for post-surgical pain management due to their effectiveness; surgical patients are four times as likely to be discharged with an opioid prescription than nonsurgical patients \citep{hinther_chronic_2019}. However, there  exists a high risk for  persistent opioid usage, which can lead to opioid misuse or overdose.  Analogously, overtreatment harms occur in  conditions such as kidney disease, where overhydration increases mortality and technique failure \citep{shu2018effect, dharmarajan2017state}.

A review of 42 large scale studies found that, among surgical patients that had not been exposed to opioids prior to surgery (``opioid-naïve'' patients), 10.4\%, 7.7\%, and 9.1\% developed chronic postoperative opioid use at 3, 6, and 12 months post-surgery, respectively \citep{hinther_chronic_2019}. Duration of opioid usage was linked to risk of misuse, with significant increases in risk of misuse with each additional prescription refill \citep{brat_postsurgical_2018}. 
Thus, it is necessary for prescribers to balance using opioids to effectively manage patients' pain post-surgery and also ensuring that patients do not experience harms from these drugs. Understanding post-surgical opioid prescription refill behavior is crucial to finding this balance between improving postoperative pain management and reducing the risk of opioid misuse \citep{brummett_new_2017}. 
Prior research on opioid refill behavior has almost exclusively been limited to studying the number of refill events and rates of refill measured at fixed time points after surgery \citep{tollemar_association_2023, brat_postsurgical_2018, clarke_rates_2014}. Studying the times between prescription refill events may provide additional insights into patient behavior, as shorter times between refill events indicate more frequent refills for a patient. However, these prescription refill ``gap'' times currently remain understudied. 
Furthermore, since opioid prescriptions can be modified to meet patients’ changing needs, incorporating the dosage level into refill analyses offers a more comprehensive understanding of pain management \citep{wetzel_opioid_2021, hill_wide_2017}. Higher opioid dosage levels have been linked to higher risk of addiction, making it all the more important for providers to identify the optimal dosage levels for their individual patients \citep{huffman_nonopioid_2015}. Additionally,  patients with different characteristics may have different risks of developing opioid misuse as well as different needs in terms of pain management, and these different outcomes may further depend on the dosage level of opioid prescriptions as well \citep{musich_prevalence_2019, young_postsurgical_2021}. Thus, it is crucial to explore the potential dosage-dependent relationships between patient characteristics and post-surgical opioid usage. Characteristics measured on a continuous scale that can be linked to the event recurrence, such as opioid refill dosage, are commonly termed “mark” variables \citep{sun_proportional_2009, gilbert_tests_2004, gilbert_2-sample_2008}. 
In our motivating post-surgical study of 11,492 patients from a major surgical and prescription record dataset, identifying patient and clinical factors for opioid refills at distinct dosage levels is of major interest. 

%\noindent \textbf{Existing Methods}
Recurrent events analysis methods typically work on one of two time scales \citep{amorim_modelling_2015}: the total time scale, where the time from study entry until each event occurrence is the outcome of interest \citep{prentice_regression_1981, andersen_coxs_1982}, or the gap time scale, where the time at risk is reset to zero for each subject after each recurrence they experience and the outcome of interest is time between recurrences \citep{prentice_regression_1981, schaubel_regression_2004, huang_marginal_2003}. 
Our work focuses on the gap time scale, as clinicians in our motivating study are primarily interested in identifying factors associated with higher hazards of prescription refills. In this context, prescription refill events align with the gap time scale, where shorter gaps between refills indicate higher refill hazards. 
A number of methods for gap time analysis have previously been developed. \citet{prentice_regression_1981} modeled the hazards of a given recurrence using an extended  Cox regression model by assuming conditional independence of gap times and conditioning on the process and covariate history; \citet{schaubel_regression_2004} proposed an estimating equations-based approach that does not require the specification of the dependence structure of gap times within subjects, and estimated the hazards of experiencing a given recurrence conditional on the previous events; \citet{huang_marginal_2003} treated a subset of recurrent events data as clustered survival data with an informative cluster size and reformulated the score function as a functional of empirical processes averaged over subjects and events within subjects to estimate hazards of recurrence. Other authors built off of \citet{huang_marginal_2003} by developing a modified within-cluster resampling approach to improve efficiency \citep{luo_analysis_2011} and incorporating the inverse cluster size weights directly into the partial likelihood contributions from each subject \citep{darlington_event-weighted_2013}. 
However, while many models for the time between recurrent events (“gap time” models) have been developed, none are able to accommodate continuous mark variables linked to the event occurrences.  % there are limited recurrent events modeling approaches that can also incorporate the additional information about event recurrences contained in the mark variable.  
Meanwhile, \citet{sun_proportional_2009} developed a mark-specific proportional hazards model for univariate survival data to estimate the hazards of event occurrence at specified mark values of interest by using kernel function-based weights. However, this method cannot address the issues that are commonly associated with gap time analysis, such as dependencies among gap times within subjects, non-representativeness of censored versus uncensored gap times after the first observed gap time, and the induced informative censoring that is intrinsic to gap time analysis. Furthermore, different gap times from the same subject can have different corresponding marks.

%\noindent \textbf{Our New Method and Paper Organization}

%Existing methods are capable of addressing the modeling of recurrent events on the gap time scale or modeling hazards for univariate failure time data at specific marks, but no approaches currently exist to model mark-specific hazards of recurrent events. 
To address these issues, we propose a new mark-specific proportional hazards model for gap times, which can model the association between the hazard of event recurrence and covariates at any chosen mark. To ensure the regression coefficients have population-average interpretations, we adopt a marginal likelihood approach, incorporating a dual weighting scheme that simultaneously accounts for the informative gap times experienced by each subject and addresses nonparametric estimation by weighting event contributions based on their proximity to the mark value of interest. Finally, the concept of mark-specific hazard models
%, along with the model proposed by \citet{sun_proportional_2009},  
extends cause-specific hazard models \citep{prentice_analysis_1978} from discrete causes to allow the cause of interest (the ``mark'') to exist on a continuous scale. 

The paper is structured as follows. In Section 2, we introduce the mark-specific proportional hazards model for recurrent events. We propose a dual weighted partial likelihood function for estimation in Section 3, and present
asymptotic results and a sandwich variance estimator for robust inference in Section 4. We explore the finite-sample performance of our method via simulations in Section 5. In Section 6, we apply our method to analyze the opioid prescription refill data. A discussion and concluding remarks are provided in Section 7. The proofs are deferred to the Supplementary Materials.

\section{Methodology} \label{sec:meth}
\subsection{Mark-specific Proportional Hazards Models for Recurrent Events}
%\noindent \textbf{Notation}
Consider a sample of $n$ independent patients.
%Let $i$ index subjects ($i=1,...,n)$ and $m_i$ represent the number of gap times the $i^{th}$ subject has. 
For the $i^{th}$ subject ($i=1, \ldots, n)$, let $j=0,1,2, \ldots$ 
index the event number ($j=0$ marks the beginning of the follow-up period). 
Starting with $j=1$, we use $T_{i(j)}$  to represent the gap time between event $j-1$ and $j$ (i.e., $j^{th}$ gap failure time) and $\bm{T}_{i}=\{ T_{i(j)}:j=1,2,\ldots \}$ to represent the recurrent event sequence. We use $C_i$ to denote the follow-up time, and  $M_i$ to represent the event index such that $\sum_{j=1}^{M_i}{T_{i(j)}} \leq C_i$ and $\sum_{j=1}^{M_i+1}{T_{i(j)}}>C_i$ (i.e., $M_i$ events are observed at or before time $C_i$ and the $M_i+1^{th}$ event is unobserved and happens after $C_i$). Note that  $M_i$ can be 0 if no events occur before the end of the follow-up period; furthermore, let $\sum_{j=1}^{0}{T_{i(j)}} \equiv 0$. Let $T_{i(M_i+1)}^+ \equiv C_i - \sum_{j=1}^{M_i}{T_{i(j)}}$ be the time at which the $M_i+1^{th}$ event is censored (on the gap time scale). Additionally, let $R_i \equiv \max (M_i, 1)$. We use $\bm{V}_i = \{V_{i(j)}:j=1,2, \ldots, M_i\}$ to denote the vector of marks associate with the $i^{th}$ person's $M_i$ observed gap times, with $V_{i(j)}$ representing the mark value associated with the $i^{th}$ person's $j^{th}$ event. Note that $V_{i(j)}$ is undefined if the $i^{th}$ subject's $j^{th}$ event is not observed. We assume marks have a known and bounded support, which we take to be [0,1] without loss of generality (rescaling the mark variable if necessary). Finally, $\bm{Z}_i$ is the time-independent covariate vector, though it is straightforward to extend to allow time-dependent covariates. 

%\pink{need to add/modify assumptions here:} 
For a given mark $v$, we assume that the bivariate recurrent event process $T_{i(j)},V_{i(j)}, j=1,\ldots R_i$ within a subject $i$ is identically distributed conditional on covariates $\bm{Z}_i$ and a subject-specific unobserved, latent variable.
%,R_i?,C_i?$
%\pink{(gamma is subject-specific frailty) - from Luo/Huang 2011}
%\pink{Luo/Huang also assumes that given $Z_i$, the hazard fn of $T_{i(j)}$ satisfies the proposed PH model...}
We also assume that censoring time $C_i$ is  independent of gap times $\bm{T}_{i}$, marks $\bm{V}_i$, and the subject-specific latent variable, conditional on $\bm{Z}_i$. 

For modeling the mark-specific hazards of event recurrence, we define the hazard of the $i^{th}$ subject experiencing the $j^{th}$ event with corresponding mark $v$ at time $t$,  given $\bm{Z}_i$, as 
\begin{equation} \label{mshazardfn}
    \lambda(t,v|\bm{Z}_i) =
    \lambda_{(j)}(t,v|\bm{Z}_i) = 
    \lim_{\Delta_1, \Delta_2 \rightarrow 0^+}
    \frac{\Pr(T_{i(j)} \in [ t,t+\Delta_1),V_{i(j)} \in [ v, v+\Delta_2) | T_{i(j)} \geq t, \bm{Z}_i  )}
    {\Delta_1 \Delta_2},
\end{equation}
which can be interpreted as the instantaneous rate of the $j^{th}$ failure with corresponding mark $v$ at time $t$, given that the $j^{th}$ failure (with any mark) did not occur prior to time $t$. Since we assume an identically distributed bivariate process for failure times and marks within subjects conditional on their covariates and subject-specific latent variables,
%and subject-specific random effect
 the hazard is not episode-specific, so  $\lambda_{(j)}(t,v|\bm{Z}_i)$ is the same across all $j$ for each subject. 
 %For our model, we assume that gap times within subjects are identically distributed (but not necessarily independent). 
%That is, conditional on covariates and subject-specific random effects, gap times are identically distributed. 
%(i.e.,  This implies that the observed complete gap times ${T{i(j)}, j = 1, \ldots, M_{i}-1})$ are identically distributed given $(C_i, M_i, T_{i(M_i+1)}^+, M_i))$.

We  propose a mark-specific, recurrent events proportional hazards model:
\begin{equation} \label{mshazardmodel}
     \lambda(t,v|\bm{Z}_i) = \lambda_0(t,v)\exp(\bm{\beta}(v)^\mathrm{T}\bm{Z}_i),
\end{equation}
where $v$ is the chosen mark of interest, $\bm{Z}_i$ is the $p \times 1$ vector of covariates for the $i^{th}$ subject,  $\lambda_0(t,v)$ is the baseline hazard function corresponding to the chosen mark of interest $v$, and $\bm{\beta}(v)$ represents the $p \times 1$ vector of regression coefficients corresponding to the fixed mark of interest $v$. 
%From this model, 
More specifically, the $q^{th}$ element of $\bm{\beta}(v)$ represents the log hazard ratio for an event recurrence corresponding to mark $v$ for a subject with the $q^{th}$ element of their covariate vector being $Z_{iq}$ versus a subject with the $q^{th}$ element of their covariate vector being the baseline covariate value, adjusting for the other covariates and irrespective of episode number.

The mark-specific hazard  (\ref{mshazardfn}) is connected to the  overall (or any-cause) hazard $\lambda_{AC}(t|\bm{Z}_i)$ via
%\begin{equation} \label{marginal hazard}
 $\lambda_{AC}(t|\bm{Z}_i)
= \int_0^\infty\lambda(t,v|\bm{Z}_i)dv$, further implying
%which gives the overall survival function and marginal density function of gap time:
%\begin{equation} \label{overall surv}
 %   S(t|\bm{Z}_i) = \exp(- \int_0^t\lambda(s|\bm{Z}_i)ds), \quad 
%f(t|\bm{Z}_i) = -\frac{d}{dt} S(t|\bm{Z}_i)
%\end{equation}
that the density of $V_{i(j)}$ given $T_{i(j)}, \bm{Z}_i$ is
\begin{equation} \label{cond mark pdf}
    f(v| T_{i(j)}=t, \bm{Z}_i) = \frac{ \lambda(t, v| \bm{Z}_i) S(t | \bm{Z}_i)}{f(t| \bm{Z}_i)}.
\end{equation}
%\noindent \textbf{Connections to Other Models}
Model (\ref{mshazardmodel}) is general, encompassing a variety of existing models \citep{sun_proportional_2009, huang_marginal_2003, luo_analysis_2011, darlington_event-weighted_2013} as special cases. For example, when $R_i = 1$ for all subjects, the model reduces to the mark-specific proportional hazards model proposed by \citet{sun_proportional_2009} for non-recurrent time-to-event data with marks;  when $\bm{\beta}(v) = \bm{\beta}$ (i.e., each parameter is constant across mark values), our model reduces to the non-mark-specific recurrent events model proposed by \citet{huang_marginal_2003} and refined by \citet{luo_analysis_2011} and \citet{darlington_event-weighted_2013}, in which recurrent events are treated as clustered survival data with informative cluster sizes.

\subsection{Dual Weighted Partial Likelihood for Estimation}

Modeling recurrent events on the gap time scale with no data modifications induces dependent censoring for events after the first observed event (i.e., for $j \geq 2$); furthermore, gap times within subjects may not be representative of each subjects' gap event time distribution \citep{huang_marginal_2003}. To remedy these two issues and following \citet{huang_marginal_2003}, we propose to drop censored gap times for subjects who have at least one observed event.  More formally, let $\delta_{i(j)} \equiv \mathds{I}(\sum_{k=1}^{j}{T_{i(j)}} \leq C_i)$ (indicator for whether the $j^{th}$ event for $i^{th}$ subject is observed or censored), and $X_{i(j)} \equiv T_{i(j)}$ if $\delta_{i(j)}=1$, or $T_{i(j)}^+$ if $\delta_{i(j)}=0$.
% $X_{i(j)} \equiv \begin{cases}
%     T_{i(j)}, \delta_{i(j)}=1\\
%     T_{i(j)}^+, \delta_{i(j)}=0
%     \end{cases}$.
Then, recalling that $R_i \equiv \max (M_i, 1)$, the analytical data is given by 
$\{ X_{i(j)}, \delta_{i(j)}, V_{i(j)}, \bm{Z}_{i} \}$, $i=1,2,\ldots n; j=1,2,\ldots, R_i$. 
%Note that if \( M_i > 0 \), then \( \delta_{i(j)} = 1 \) for all \( j \), meaning all episodes for subject \( i \) are uncensored in the analytical dataset. 
For subject $i$, if \( R_i \geq 2 \), this implies \( M_i \geq 2 \); in this case, all refill episodes for that subject are observed without censoring, and we thus define the \emph{marked counting process}
\(N_{i(j)}(t, v) = \mathds{I}(T_{i(j)} \leq t, V_{i(j)} \leq v),\) and the \emph{at-risk process} \(Y_{i(j)}(t) = \mathds{I}(T_{i(j)} \geq t),\) both of which are fully observable in the dataset for \( j = 1, \ldots, M_i \).
On the other hand, if \( R_i = 1 \), then \( M_i \in \{0, 1\} \), allowing for the presence of censoring. In this case, we define the \emph{observed marked counting process} as \(
N_{i(j)}(t, v) = \mathds{I}(X_{i(j)} \leq t, \delta_{i(j)} = 1, V_{i(j)} \leq v),\)
and the \emph{at-risk process} as
\(Y_{i(j)}(t) = \mathds{I}(X_{i(j)} \geq t).\)
%Note that if $M_i > 0$, then $\delta_{i(j)}=1$ for all $j$ for subject $i$ in our analytical dataset.  We  introduce $N_{i(j)}(t,v)= \mathds{I}(X_{i(j)} \leq t, \delta_{i(j)} = 1, V_{i(j)} \leq v)$,  the observed {\em marked} counting process, and   $Y_{i(j)}(t) = \mathds{I}(X_{i(j)} \geq t)$, the at-risk process. 

When constructing a likelihood function, we introduce a dual weighting scheme that serves two distinct purposes.  First, to account for the informative nature of the number of recurrent events per subject, we weight the contributions from each subject by $\omega_i = 1/R_i$; events from subjects with more observed recurrences are downweighted to avoid giving events from these subjects too much influence. Second, for nonparametric estimation of $\bm{\beta}(v)$, we weight the contributions for each observed event depending on their corresponding mark \citep{sun_proportional_2009}. We use a  kernel function to allow events with marks closer to the mark of interest to count more heavily in the estimation procedure and vice versa.

The resulting event-weighted, recurrent event localized log partial pseudo-likelihood for $\bm{\beta}(v)$  at a fixed mark $v$ is
\begin{equation}
\begin{aligned}
    l(v, \bm{\beta}(v)) = 
    \sum_{i=1}^{n} \Bigg [ \omega_i
    \sum_{j=1}^{R_i}
    \bigg\{
    \int_{0}^{1}
    \int_{0}^{\tau}
     K_h(u-v)
     \bigg [
     \bm{\beta}(v)^\mathrm{T}\bm{Z}_i - 
    \log \big [ \sum_{k=1}^{n} \omega_k \sum_{l=1}^{R_k} Y_{k(l)}(t) \exp (\bm{\beta}(v)^\mathrm{T} \bm{Z}_k )\big ] \bigg] \\
     \times N_{i(j)}(dt, du)
     \bigg \}
     \Bigg ],
\end{aligned}
\end{equation}
where $\omega_i=1/R_i$ and 
$K_h(x) =\frac{1}{h} K(\frac{x}{h})$ is a kernel function with a bandwidth $h$, where $K(x)$ is a symmetric and continuous kernel function with bounded compact; for example, the Epanechnikov kernel $K(x) =\frac{3}{4}(1-x^2)$ for $ |x|<1$. 
%\frac{\partial l(v,\bm{\beta}(v))}{\partial (\bm{\beta}(v))}

The corresponding pseudo-score function is
\begin{equation} \label{eq4}
\begin{aligned}
    \bm{U}_{\bm{\beta}}(v, \bm{\beta}(v)) =  \nabla_{\bm{\beta}} l(v,\bm{\beta}(v))= 
    \sum_{i=1}^{n} \bigg [ \omega_i
    \sum_{j=1}^{R_i}
    \bigg\{
    \int_{0}^{1}
    \int_{0}^{\tau}
     K_h(u-v)
    \bigg [
    \bm{Z}_i - 
    \frac{S^{(1)}(t,\bm{\beta}(v))}
        {S^{(0)}(t,\bm{\beta}(v))}
    \bigg ]
    N_{i(j)}(dt,du)
    \bigg \} \bigg ],
\end{aligned}    
\end{equation}
where $S^{(q)}(t,\bm{\beta}(v)) = n^{-1} \sum_{i=1}^{n} [ \omega_i \sum_{j=1}^{R_i} {Y_{i(j)} (t) \exp(\bm{\beta}(v)^\mathrm{T} \bm{Z}_i ) \bm{Z}_i}^{\otimes q} ], q=0,1,2$. 
Solving $  \bm{U}_{\bm{\beta}}(v, \bm{\beta}(v)) = 0$ yields $\hat{\bm{\beta}}(v)$, the maximum pseudo-partial likelihood estimator of $\bm{\beta}(v)$, which can be computed using the Newton-Raphson algorithm.

We also introduce the second derivative of the log partial pseudo-likelihood, which will be used for inference and asymptotic results in the next section:
%  \frac{\partial^2 l(v,\bm{\beta}(v))}{\partial (\bm{\beta}(v))^2}
\begin{equation}\label{eq5}
   \nabla^2_{\bm{\beta} \bm{\beta}} l(v,\bm{\beta}(v)) = -\sum_{i=1}^{n} \bigg [
    \omega_i 
    \sum_{j=1}^{R_i}
    \bigg\{
    \int_{0}^{1}
    \int_{0}^{\tau}
     K_h(u-v)
    \bm{J}_n(t,\bm{\beta}(v))
    N_{i(j)}(dt,du)
    \bigg \} \bigg ],
\end{equation}
where $\bm{J}_n(t,\bm{\beta}(v)) = \frac{S^{(2)}(t,\bm{\beta}(v))} {S^{(0)}(t,\bm{\beta}(v))} - \bigg[ \frac{S^{(1)}(t,\bm{\beta}(v))} {S^{(0)}(t,\bm{\beta}(v))} \bigg]^{\otimes 2}$.

\subsection{Asymptotic Results and Statistical Inference}
We establish the consistency and asymptotic normality of $\hat{\bm{\beta}}(v)$ at interior values of the mark $v \in [a,b] \subset (0,1)$.
We list the required regularity conditions.

\noindent\textbf{Regularity Conditions}

\noindent Let  $s^{(q)}(t,\bm{\theta} )= \EX [ S^{(q)}(t,\bm{\theta})]$ for $q=0,1,2$ and $\bm{J}(t,\bm{\theta}) = \frac{s^{(2)}(t,\bm{\theta})} {s^{(0)}(t,\bm{\theta})} - \Big[ \frac{s^{(1)}(t,\bm{\theta})}
        {s^{(0)}(t,\bm{\theta})} \Big]^{\otimes 2}$.

 \begin{itemize}
     \item[\textbf{(C1)}] We assume that $\bm{\beta}(v)$ has component-wise continuous second derivatives on $[0,1]$. The second partial derivative of $\lambda_0(t,v)$ with respect to $v$ exists and is continuous on $[0,\tau] \times [0,1]$.  $\bm{Z}$ satisfies $\EX[||\bm{Z}||^4 \exp(2B||\bm{Z}||)] < \infty$, where $B$ is a constant such that $(v,\bm{\beta}(v)) \in [0,1] \times (-B, B)^p$.
     \item[\textbf{(C2)}] For $q=0,1,2$, each component of $s^{(q)}(t,\bm{\theta})$ is continuous on $[0,\tau] \times [-B,B]^p$.
     \item[\textbf{(C3)}] \sloppy It is assumed that $s^{(0)}(t,\bm{\theta}) > 0$ on $[0,\tau] \times [-B,B]^p$ and the matrix $\bm{\Sigma}(v) = \int_{0}^{\tau} {\bm{J}(t,\bm{\beta}(v)) \times \lambda_0(t,v) s^{(0)}(t,\bm{\beta}(v)) dt} $ is positive definite.
     %\item[\textbf{(C4)}] We assume that $\EX [ N_{i(j)}(dt, dv)|\mathcal{F}_{i,j, t-}] = \EX [ N_{i(j)}(dt, dv) | Y_{i(j)}(t), \bm{Z}_i]$, where $\mathcal{F}_{i,j,t}$ is the right-continuous filtration generated by $ \{N_{i(j)}(s,v), Y_{i(j)}(s), 0\leq s\leq t; \bm{Z}_i\}$. \eycm{this may not be needed(?)}
     \item[\textbf{(C4)}] The kernel function $K(.)$ is symmetric with support $[-1,1]$ and of bounded variation. The bandwidth satisfies $nh^3 \rightarrow \infty$ and $nh^5 \rightarrow 0$ as $n \rightarrow \infty$.
 \end{itemize}

\sloppy Conditions \textbf{(C1)}--\textbf{(C3)} are standard  for survival analysis and used throughout the proofs; under \textbf{(C1)} and the Donsker property, we also have that $\sup_{t \in [0,\tau], \bm{\theta} \in [-M,M]^p} ||S^{(q)}(t,\bm{\theta}) - s^{(q)}(t,\bm{\theta})|| = O_p(n^{-1/2})$.
\textbf{(C4)} provides conditions for valid kernel functions and bandwidths for our method.
Under our defined mark-specific recurrent events hazard function (\ref{mshazardfn}), we have %$\EX[N_{i(j)} (dt,dv)|\mathcal F_{i,j,t}-] = $
$\EX[N_{i(j)} (dt,dv)|Y_{i(j)}(t), \bm{Z}_i] = Y_{i(j)}(t)\lambda(t,v|\bm{Z}_i) dt dv$. 
Thus, we can define $M_{i(j)}(t,v) = \int_{0}^{t} { \int_{0}^{v} {[N_{i(j)} (ds, dx) - Y_{i(j)}(s)\lambda(s,x|\bm{Z}_i) ds dx ]}}$.

%\subsection{Asymptotic Results}

\noindent \textbf{Theorem 1 (Uniform Consistency):} Under \textbf{(C1)}--\textbf{(C4)},% for fixed $0<a<b<1$,   
\begin{equation}
    {\sup_{v \in [a,b]} |\hat{\bm{\beta}}(v)-  \bm{\beta}(v)| = o_p(1)}.
\end{equation}
\noindent \textbf{Theorem 2 (Asymptotic Normality):} Under \textbf{(C1)}--\textbf{(C4)}, for any $v \in [a,b]$, 
\begin{equation}
    \sqrt{nh}(\hat{\bm{\beta}}(v)-\bm{\beta}(v)) \xrightarrow{d} N(0, \bm{\Gamma}(v)),
\end{equation}
with $
\bm{\Gamma}(v) = \bm{\Sigma}^{-1}(v)   \bm{\Lambda}(v) \bm{\Sigma}^{-1}(v)$,
where $\bm{\Sigma}^{-1}(v)$ is the pseudo-information matrix  and $\bm{\Lambda}(v)$ is the asymptotic variance of the pseudo-score function (formally defined in the proof of \textbf{Theorem 2}).

%[\Big]{\frac{\partial^2 l(v,\bm{\beta}(v))}{\partial (\bm{\beta}(v))^2}}
Detailed proofs for \textbf{Theorems 1 and 2} are in the Supplementary Materials. The asymptotic normality results lay the ground for approximate inference based on normal distributions. In practice, to estimate $\bm{\Gamma}(v)$, we can use estimated versions of each component of $\bm{\Gamma}(v)$. Specifically, let $\hat{\bm{\Gamma}}(v) = \hat{\bm{\Sigma}}^{-1}(v)  \hat{\bm{\Lambda}}(v)\hat{\bm{\Sigma}}^{-1}(v)$
where $\hat{\bm{\Sigma}}(v) = -\frac{1}{n} \Big[ \evalat {\nabla^2_{\bm{\beta} \bm{\beta}} l(v,\bm{\beta}(v))} {\hat{\bm{\beta}}(v)} \Big]$ and $\hat{\bm{\Lambda}}(v)$ is the estimated variance of the pseudo-score function, which can be found using the pseudo-score residuals. Let $\hat{\bm{w}}_{i(j)}(v)$ represent the pseudo-score residual for the $i^{th}$ subject's $j^{th}$ episode corresponding to mark $v$: $\hat{\bm{w}}_{i(j)}(v) = \omega_i \int_{0}^{\tau}{
    \big\{
     \delta_{i(j)} K_h(V_{i(j)}-v)
    \big [
    \bm{Z}_i - 
    \frac{S^{(1)}(s,\hat{\bm{\beta}}(v))}
        {S^{(0)}(s,\hat{\bm{\beta}}(v))}
    \big ]
    \big \}}d\hat{M}_{i(j)}(s)$. Then $\hat{\bm{\Lambda}}(v) = \frac{1}{n} 
\sum_{i=1}^{n} \big [
\sum_{j=1}^{R_i} 
\sum_{k=1}^{R_i}
{\hat{\bm{w}}_{i(j)}}(v)
{\hat{\bm{w}}_{i(k)}}(v)^\mathrm{T} \big ]
$.

\subsection{Model Implementation}

In practical implementation, we must specify a bandwidth \( h \). When the mark distribution is relatively uniform and the goal is to study mark-specific (MS) hazards over a broad range, a common bandwidth can be used. We adopt a data-adaptive method inspired by \citet{cao_regression_2015} to select a uniform bandwidth by minimizing the integrated mean squared error (IMSE) across marks, without relying on cross-validation.  

Specifically, we evaluate a prespecified set of candidate bandwidths for each mark $v$ in the set of marks of interest. 
For each mark $v$ in the set of marks of interest, we first use the full analytical dataset to find the $\hat{\bm{\beta}}(v)$ that would be estimated using each choice of $h$ in the prespecified set of candidate bandwidths (call these $\hat{\bm{\beta}}_h(v)$). 
Next, we regress the estimated $\hat{\bm{\beta}}_h(v)$ on $h^2$ in the prespecified range of candidate $h$ values to estimate a slope value $\hat{C}(v)$. 
We also estimate the variance of $\hat{\bm{\beta}}_h(v)$ by splitting the analytical dataset randomly into two halves, obtaining coefficient estimates $\hat{\bm{\beta}}_{h1}(v)$ and $\hat{\bm{\beta}}_{h2}(v)$ for each half-sample, and calculating $\hat{V}_h(v) = (\hat{\bm{\beta}}_{h1}(v) - \hat{\bm{\beta}}_{h2}(v))^2 / 4$. 
The mark-specific mean squared error (MSE) for a particular candidate $h$ is then calculated as $MSE_h(v) =\{\hat{C}(v)\}^2h^4+\hat{V}_h(v)$, based on \textbf{Theorem 1} of \citet{cao_regression_2015}. 
Then, the IMSE for each candidate $h$ is taken as the average of the calculated mark-specific mean squared errors across the marks of interest for a particular candidate $h$. The candidate $h$ that achieves the lowest IMSE is chosen as the uniform optimal bandwidth and used as the common bandwidth for all the mark-specific hazards models.

However, when the mark distribution is highly skewed or only a limited set of mark values is of interest, it may be necessary to select different bandwidths for each mark. In such cases, we recommend evaluating each mark individually to choose a bandwidth that balances two competing goals: ensuring sufficient data around the mark for stable model fitting, while maintaining enough localization to yield meaningful mark-specific estimates.

\section{Simulation Study}

%\subsection{Simulations for Linear $\beta(v)$}
We performed a simulation study to evaluate the finite sample performance of the proposed model and compare it against an existing non-mark-specific hazards model for recurrent events \citep{huang_marginal_2003, luo_analysis_2011, darlington_event-weighted_2013}. 
Mimicking our real data example, we generated $J=5$ recurrent event gap times and corresponding marks for $n=1000$ subjects for each simulation setting. 
We studied two settings corresponding to different forms of the true mark-specific parameter $\beta(v)$; specifically, we generated mark-specific recurrent events data from the underlying model $\lambda(t,v|Z)= \lambda_0(t,v)\exp((\beta_1+\beta_2 v)Z)$ to explore model performance when the true mark-specific parameter $\beta(v)$ has a linear form,  and from the underlying model $\lambda(t,v|Z) = \lambda_0(t,v) \exp((\beta_1+\beta_2v^2)Z)$ to explore performance when $\beta(v)$ has a quadratic form. 

Here, $\lambda_0(t,v)$ is the baseline mark-specific hazard function, which controls the dependence between gap times and marks in the data generation; we set $\lambda_0(t,v)=\exp(\beta_0 v)(t+v)$, which leads to $T \text{ and } V$ being dependently generated. 
Chosen values for $\beta_0,\beta_1, \beta_2$ represent chosen parameters that determine the underlying mark-specific hazard ratio, with $\beta_1+\beta_2 v$ being the mark-specific parameter of interest $\beta(v)$ in the linear $\beta(v)$ setting and $\beta_1+\beta_2v^2$ being the mark-specific parameter of interest in the quadratic $\beta(v)$ setting. The values of these parameters can also impact the underlying distribution of marks in the generated datasets. 
For each setting (linear $\beta(v)$ and quadratic $\beta(v)$), we considered two choices of parameter values: $(\beta_0, \beta_1, \beta_2)=(0.3,-0.5,0.5)$, which corresponds to $\beta(v) = -0.5+0.5v$ in the linear setting and $\beta(v) = -0.5+0.5v^2$ in the quadratic setting (hazard ratio increases across marks and generated marks have a left-skewed distribution), and $(\beta_0, \beta_1, \beta_2)=(0.3,-0.5,-1.5)$, corresponding to $\beta(v)=-0.5-1.5v$ in the linear setting and $\beta(v) = -0.5-1.5v^2$ (hazard ratio decreases across marks and generated marks have a relatively uniform distribution). 
The subject-specific covariate $Z$ is generated from the $Bern(0.5)$ distribution. 

Gap times that were correlated within a subject were generated using an inverse-transform sampling procedure based on the overall survival function. First, for each subject, one subject-specific variable $A_i$ from a $N(0,v)$ distribution and $J$ within-subject, episode-specific variables $B_{i(j)}$ from a $N(0,1-v)$ distribution were generated, where $v \in (0,1)$. We set $v=0.25$. Note that $A_i+B_{i(j)}$ now follows a $N(0,1)$ distribution. Then, we take $1-\Phi (A_i+B_{i(j)})$ to act as $S(T_{i(j)}|Z_i)$. Gap event times $T_{i(j)}$ were then back-solved from the overall survival function $S(T_{i(j)}|Z_i)$ (derived from the overall hazard function) with the previously generated $Z_i$ plugged in. Within-subject correlated marks were generated via a similar inverse transform sampling approach by back-solving  $F(V_{i(j)}|T_{i(j)}, Z_i)$   [see (\ref{cond mark pdf})] using the previously generated $T_{i(j)}$ and $Z_i$ values. A censoring time on the total time scale was generated for each subject from a $Unif(0, \tau_c)$ distribution, and $\tau_c$ was chosen for each simulation setting to ensure around 25\% of the gap times were censored.

Ultimately, our main simulation study had two different data generation settings, corresponding to the combinations of the two mark-specific parameter forms (linear or quadratic $\beta(v)$) and the two mark-specific parameter values ($(\beta_0, \beta_1, \beta_2)=(0.3,0.5, -0.5)$ or $(0.3,0.5,-1.5)$). For each simulation setting, we compared the performance of our proposed mark-specific model and the existing non-mark-specific model at mark values of interest ranging between $0.1$ to $0.9$, incremented by $0.1$. The bandwidth used for each tested simulation setting depended on the underlying distribution of marks in the generated data from that setting (see the note on bandwidth selection at the end of our Methods section for more details). Specifically, for settings with $(\beta_0,\beta_1,\beta_2)=(0.3,0.5,-0.5)$, we chose different bandwidths that were specific to each mark of interest as the underlying distribution of marks in the generated datasets were left-skewed; these separate bandwidth values for each mark were chosen such that around 1000 observed events around a given mark of interest were captured within the bandwidth. 
For settings with $(\beta_0,\beta_1,\beta_2)=(0.3,0.5,-1.5)$, we chose a common bandwidth to use across marks for each replication of simulations with $\beta(v)=-0.5+0.5v$, as the datasets generated with this $\beta(v)$ choice produced a relatively uniform mark distribution. For the uniform optimal bandwidth selection procedure here, we tested a range of $h$ values between $0.05 \text{ and } 0.80$, incremented by $0.01$.
We report the average bias and coverage probability of parameter estimates found using the two competing methods at mark values of 0.3, 0.5, and 0.7 for each simulation setting and modeling method across 1000 replications in Table \ref{tab:simres_biasCP}. Table \ref{tab:simres_SDSE} contains the average standard errors and empirical standard deviations of parameter estimates across 1000 replications from our proposed method. 

\begin{table}[ht]
    \caption{Comparison of average bias and coverage probabilities across 1000 replications for the proposed mark-specific hazards model for recurrent events and the existing non-mark-specific hazards model for recurrent events across simulation settings}
    \label{tab:simres_biasCP}
    \centering
    \begin{adjustbox}{width=1\textwidth}
    \begin{tabular}{cccllllllllll}
    \hline
 & & & \multicolumn{10}{c}{\textbf{Mark}}\\
 \cline{4-13}

         &  &  &  \multicolumn{2}{c}{\textbf{0.1}}&  \multicolumn{2}{c}{\textbf{0.3}}&  \multicolumn{2}{c}{\textbf{0.5}}&    \multicolumn{2}{c}{\textbf{0.7}}&\multicolumn{2}{c}{\textbf{0.9}}\\
        \cline{4-13}

         \multirow[c]{2}{*}{\textbf{Form of $\beta(v)$}}& \multirow[c]{2}{*}{\textbf{$(\beta_0,\beta_1,\beta_2)$}}&\multirow[c]{2}{*}{\textbf{Method}}& \multirow{2}{1em}{\textbf{Avg. Bias}}& \multirow{2}{1em}{\textbf{Cov. Prob.}}& \multirow{2}{1em}{\textbf{Avg. Bias}}& \multirow{2}{1em}{\textbf{Cov. Prob.}}& \multirow{2}{1em}{\textbf{Avg. Bias}}& \multirow{2}{1em}{\textbf{Cov. Prob.}}& \multirow{2}{1em}{\textbf{Avg. Bias}}& \multirow{2}{1em}{\textbf{Cov. Prob.}}& \multirow{2}{1em}{\textbf{Avg. Bias}}& \multirow{2}{1em}{\textbf{Cov. Prob.}}\\
 \\
 \hline 
          Linear&  (0.3, -0.5, 0.5)& MS (mark-specific $h$)&   0.037&  0.950&  0.002&  0.956& 0.006&  0.946&   0.009& 0.947& -0.002& 0.946
\\
         &  &  Non-MS
&  0.262&  0.047&  0.162&  0.364&  0.062&  0.841&    -0.038&0.907&-0.138&0.506
\\
         &  (0.3, -0.5, -1.5)&  MS (uniform $h$)&  -0.106&  0.870&  -0.020&  0.937&  -0.015&  0.954&    -0.009&0.939&0.083&0.852
\\
         &  &  Non-MS
&  -0.659&  0.000&  -0.359&  0.010&  -0.059&  0.892&    0.241&0.232&0.541&0.000
\\
         Quadratic&  (0.3, -0.5, 0.5)&  MS (mark-specific $h$)&  0.006&  0.964&  -0.004&  0.949&  -0.001&  0.960&    0.003&0.963&-0.005&0.947
\\
         &  &  Non-MS
&  0.225&  0.105&  0.185&  0.261&  0.105&  0.680&    -0.015&0.944&-0.175&0.295
\\
         &  (0.3, -0.5, -1.5)&  MS (uniform $h$)&  -0.044&  0.928&  -0.043&  0.919&  -0.040&  0.933&    -0.029&0.941&0.128&0.772
\\
         &  &  Non-MS
&  -0.539&  0.000&  -0.419&  0.000&  -0.179&  0.408&    0.181&0.406&0.661&0.000
\\ \hline
    \end{tabular}
    \end{adjustbox}
\end{table}

\begin{table}[ht]
    \caption{Empirical standard deviations and average estimated standard errors across 1000 replications for parameter estimates from the proposed mark-specific hazards model for recurrent events across simulation settings}
    \centering
         \begin{adjustbox}{width=1\textwidth}
    \begin{tabular}{cccllllllllll}
    \hline
         &  &  &  \multicolumn{10}{c}{Mark}\\
          \cline{4-13}

         &  &  &  \multicolumn{2}{c}{\textbf{0.1}}&  \multicolumn{2}{c}{\textbf{0.3}}&  \multicolumn{2}{c}{\textbf{0.5}}&    \multicolumn{2}{c}{\textbf{0.7}}&\multicolumn{2}{c}{\textbf{0.9}}\\
         \cline{4-13}

         \multirow[c]{2}{*}{\textbf{Form of $\beta(v)$}}& \multirow[c]{2}{*}{\textbf{$(\beta_0,\beta_1,\beta_2)$}}&\multirow[c]{2}{*}{\textbf{Method}}& \multirow{2}{1em}{\textbf{Emp. SD}}& \multirow{2}{1em}{\textbf{Avg. SE}}& \multirow{2}{1em}{\textbf{Emp. SD}}& \multirow{2}{1em}{\textbf{Avg. SE}}& \multirow{2}{1em}{\textbf{Emp. SD}}& \multirow{2}{1em}{\textbf{Avg. SE}}& \multirow{2}{1em}{\textbf{Emp. SD}}& \multirow{2}{1em}{\textbf{Avg. SE}}& \multirow{2}{1em}{\textbf{Emp. SD}}& \multirow{2}{1em}{\textbf{Avg. SE}}\\
         \\
         \hline

         Linear&  (0.3, -0.5, 0.5)
&  MS (mark-specific $h$)&  0.155&  0.157&  0.153&  0.156&     0.156&0.156&0.157&0.155&  0.174& 0.168
\\
         &  
&  Non-MS
&  0.072&  0.070&  0.072&  0.070&     0.072&0.070&0.072&0.070&  0.072& 0.070
\\
         &  (0.3, -0.5, -1.5)
&  MS (uniform $h$)&  0.173&  0.159&  0.142&  0.130&     0.128&0.124&0.134&0.125&  0.169& 0.147
\\
         &  
&  Non-MS
&  0.092&  0.088&  0.092&  0.088&     0.092&0.088&0.092&0.088&  0.092& 0.088
\\
         Quadratic&  (0.3, -0.5, 0.5)
&  MS (mark-specific $h$)&  0.171&  0.167&  0.159&  0.159&     0.152&0.154&0.141&0.149&  0.151& 0.154
\\
         &  
&  Non-MS
&  0.071&  0.071&  0.071&  0.071&     0.071&0.071&0.071&0.071&  0.071& 0.071
\\
         &  (0.3, -0.5, -1.5)
&  MS (uniform $h$)&  0.156&  0.162&  0.131&  0.124&     0.120&0.115&0.118&0.117&  0.141& 0.165
\\
         &  
&  Non-MS
&  0.083&  0.084&  0.084&  0.083&     0.084&0.083&0.084&0.083&  0.083& 0.084
\\
\hline
    \end{tabular}
    \end{adjustbox}
    \label{tab:simres_SDSE}
\end{table}

In all simulation settings of interest, the average bias from our proposed mark-specific model appears to remain low across all marks of interest, confirming the unbiasedness of our method (Table \ref{tab:simres_biasCP}). 
Meanwhile, the average bias of parameters estimated from the non-mark-specific model is higher than average bias from the mark-specific model at every mark of interest. Note that because the non-mark-specific model cannot account for the mark-specific nature of the true parameter, the parameter estimates across the range of mark values of interest are the same and represent the average parameter estimate across all marks. Thus, our proposed method's estimation clearly outperforms that of the existing non-mark-specific method when the true parameter value differs depending on the mark.

For the settings where parameter values $(\beta_0, \beta_1, \beta_2)=(0.3,-0.5,0.5)$ were used (leading to a more skewed mark distribution and the use of the manual bandwidth selection method for choosing bandwidths), we found that our proposed mark-specific model had coverage probabilities ranging from 0.944 to 0.956 in the linear $\beta(v)$ setting, and 0.947 to 0.964 in the quadratic $\beta(v)$ setting, confirming the feasibility of using our method for inference (see Table \ref{tab:simres_biasCP}). 

In settings where the generated data had a more uniform mark distribution and the uniform optimal bandwidth was used (i.e., settings where $(\beta_0, \beta_1, \beta_2) =(0.3,-0.5,-1.5)$), coverage probabilities at the boundary mark values were lower but 95\% coverage was achieved at mark values away from the boundary values. Specifically, coverage probabilities ranged between 
0.852 (at $v=0.9$) to 0.954 (at $v=0.5$) in the linear $\beta(v)$ setting, and 0.772 (at $v=0.9$) to 0.941 (at $v=0.7$) in the quadratic $\beta(v)$ setting (see Table \ref{tab:simres_biasCP}).

In all simulation settings, our proposed method has far superior coverage compared to the non-mark-specific method. For example, for the setting with linear $\beta(v)$ and true $(\beta_0, \beta_1, \beta_2) = (0.3,-0.5,-1.5)$,  the coverage probability under the mark-specific method at mark 0.3 was 0.937, while the coverage probability under the non-mark-specific method was only 0.010. While not all differences in coverage probability between the mark-specific model and the non-mark-specific model at other marks are this extreme, the non-mark-specific model coverage probabilities are highly variable across marks in all tested settings, often with coverage probabilities equal to or close to 0 at more extreme mark values (Table \ref{tab:simres_biasCP}).
Finally, Table \ref{tab:simres_SDSE} shows that, in all tested simulation settings, the average standard error is close to the empirical standard deviation of the estimated parameters using our proposed method, confirming the utility of the proposed robust sandwich variance estimator.

\section{Analysis of the Post-surgical Opioid Prescription Refill Data }

\subsection{Data, Variables and Model Fitting}

We fit the proposed mark-specific recurrent hazards model, i.e., Model \eqref{mshazardmodel},  to study post-surgical opioid prescription refill behavior using clinical and prescription record data from 11,492 opioid-naïve patients after surgery. The dataset includes surgical patients from 2017 to 2019 who received opioid prescriptions for pain management after surgery, with time at risk beginning on the day of surgery (when they received their first prescription); the detailed patient demographics, clinical characteristics, and pre-operative prescription record information are reported in Table \ref{tab:opioid_descriptives}. The median number of post-surgical refill events experienced by patients was 2.00 (IQR: 1.00-5.00), the median gap time length between consecutive refills was 64.0 days (IQR: 13.0-434.0), and the median dosage of refill events was 112.5 Opioid Morphine Equivalents (OMEs) (IQR: 60.00-300.00).
Our analytical goal is to identify factors associated with the hazard of opioid refills at different dosage levels and explore how these associations vary across dosages. 
%Key patient characteristics (race, BMI, smoking status, cancer status) and surgical factors (approach, location) are examined. 
We also fit the existing non-mark-specific recurrent hazards model \citep{huang_marginal_2003, luo_analysis_2011, darlington_event-weighted_2013} on the same dataset to serve as a comparison to the mark-specific results.
%By understanding refill behavior across dosages, clinicians can make personalized prescription decisions to optimize pain management while reducing the risk of opioid misuse.

\begingroup
\renewcommand\arraystretch{0.5}
    \begin{longtable}{lc}
\caption{Descriptive Characteristics of Post-Surgical Opioid Prescription Refill Patients} \label{tab:opioid_descriptives} \\     
\hline
\multicolumn{1}{l}{\textbf{Characteristic}} & \multicolumn{1}{c}{\textbf{N = 11,492\(^1\)}} \\ \hline 
\endfirsthead

\multicolumn{2}{c}%
{{\itshape  \tablename\ \thetable{} -- continued from previous page}} \\
\hline \multicolumn{1}{l}{\textbf{Characteristic}} & \multicolumn{1}{c}{\textbf{N = 11,492\(^1\)}} \\ \hline 
\endhead

\hline \multicolumn{2}{r}{{Continued on next page}} \\ \hline
\endfoot

\endlastfoot

         \textbf{Sex} & \\
         \hspace{1em} Female & 6,351 (55\%) \\
         \hspace{1em} Male & 5,141 (45\%) \\
                 \textbf{Race} & \\
        \hspace{1em} White & 10,308 (90\%) \\
        \hspace{1em} Black & 1,038 (9.0\%) \\
        \hspace{1em} Asian & 99 (0.9\%) \\
        \hspace{1em} Other & 47 (0.4\%) \\
        \textbf{Age} & \\
        \hspace{1em} 18 to $<$25 & 418 (3.6\%) \\
        \hspace{1em} 25 to $<$35 & 1,034 (9.2\%) \\
        \hspace{1em} 35 to $<$45 & 1,491 (13\%) \\
        \hspace{1em} 45 to $<$55 & 2,075 (18\%) \\
        \hspace{1em} 55 to $<$65 & 2,557 (22\%) \\
        \hspace{1em} 65 to $<$75 & 2,327 (20\%) \\
        \hspace{1em} 75+ & 1,590 (14\%) \\
        \textbf{BMI} & \\
        \hspace{1em} $<$25 & 2,717 (24\%) \\
        \hspace{1em} 25 to $<$30 & 3,686 (32\%) \\
        \hspace{1em} 30 to $<$35 & 2,588 (22\%) \\
        \hspace{1em} 35 to $<$40 & 1,398 (12\%) \\
        \hspace{1em} 40 to $<$45 & 632 (5.5\%) \\
        \hspace{1em} 45 to $<$50 & 274 (2.4\%) \\
        \hspace{1em} 50+ & 200 (1.7\%) \\
        \textbf{Ascites} & 69 (0.6\%) \\
        \textbf{Cancer} & 954 (8.3\%) \\
        \textbf{Congestive Heart Failure} & 56 (0.5\%) \\
        \textbf{Chronic Condition} & 285 (2.5\%) \\
        \textbf{Chronic Obstructive Pulmonary Disease} & 552 (4.8\%) \\
        \textbf{Diabetes} & \\
        \hspace{1em} No diagnosis & 9,941 (87\%) \\
        \hspace{1em} Non-insulin & 1,064 (9.3\%) \\
        \hspace{1em} Insulin & 487 (4.2\%) \\
        \hspace{1em} Dialysis & 60 (0.5\%) \\
        \textbf{Functional Status} & \\
        \hspace{1em} Independent & 11,252 (98\%) \\
        \hspace{1em} Unknown & 30 (0.3\%) \\
        \hspace{1em} Not Independent & 210 (1.8\%) \\
        \textbf{Hypertension} & 4,821 (42\%) \\
        \textbf{Sepsis Status} & \\
        \hspace{1em} No & 10,766 (94\%) \\
        \hspace{1em} Sepsis & 493 (4.3\%) \\
        \hspace{1em} Severe Sepsis & 233 (2.0\%) \\
        \textbf{Sleep Apnea} & 2,672 (23\%) \\
        \textbf{Smoking} & 2,105 (18\%) \\
        \textbf{Ventilator} & 46 (0.4\%) \\
        \textbf{Surgical Type} & \\
        \hspace{1em} Minor Hernia & 2,940 (26\%) \\
        \hspace{1em} Abdominal Hernia & 481 (4.2\%) \\
        \textbf{Procedure} & \textbf{n (\%)} \\
        \hspace{1em}Carotid Endarterectomy & 81 (0.7\%) \\
        \hspace{1em}Creation, Re-siting, or Closure of Ileostomy or Colostomy & 109 (0.9\%) \\
        \hspace{1em}Laparoscopic Anti-Reflux and Hiatal Hernia Surgery & 146 (1.3\%) \\
        \hspace{1em}Laparoscopic Appendectomy & 1,093 (9.5\%) \\
        \hspace{1em}Laparoscopic Cholecystectomy & 2,748 (24\%) \\
        \hspace{1em}Laparoscopic Colectomy & 564 (4.9\%) \\
        \hspace{1em}Laparoscopic Hysterectomy & 809 (7.0\%) \\
        \hspace{1em}Open Appendectomy & 81 (0.7\%) \\
        \hspace{1em}Open Cholecystectomy & 79 (0.7\%) \\
        \hspace{1em}Open Colectomy & 594 (5.2\%) \\
        \hspace{1em}Open Small Bowel Resection or Enterolysis & 248 (2.2\%) \\
        \hspace{1em}Thyroidectomy & 250 (2.2\%) \\
        \hspace{1em}Total Abdominal Hysterectomy & 307 (2.7\%) \\
        \hspace{1em}Vaginal Hysterectomy & 468 (4.1\%) \\
        \hspace{1em}Other & 494 (4.3\%) \\
        \textbf{Surgical Priority} & \textbf{n (\%)} \\
        \hspace{1em}Elective & 8,164 (71\%) \\
        \hspace{1em}Urgent/Emergent & 3,328 (29\%) \\
        \textbf{Surgery Location} & \textbf{n (\%)} \\
        \hspace{1em}Outpatient & 4,922 (43\%) \\
        \hspace{1em}Inpatient & 6,570 (57\%) \\
        \textbf{Surgery Approach} & \textbf{n (\%)} \\
        \hspace{1em}Minimally Invasive & 7,578 (66\%) \\
        \hspace{1em}Open & 3,914 (34\%) \\
        \textbf{Benzodiazepines} &  \\
        \hspace{1em}No. of Preoperative Prescriptions & 0.38 (1.66) \\
        \hspace{1em}No. of Preoperative Fills & 21 (109) \\
        \hspace{1em}No. of Prescribers by Specialty & 0.12 (0.39) \\
        \hspace{1em}No. of Pharmacies & 0.12 (0.37) \\
        \hspace{1em}No. of Drug Products & 0.11 (0.34) \\
        \hspace{1em}Average Days Supply & 3 (11) \\
        \textbf{Gabapentin} &  \\
        \hspace{1em}No. of Preoperative Prescriptions & 0.0600 (0.5102) \\
        \hspace{1em}No. of Preoperative Fills & 6.7467 (61.2598) \\
        \hspace{1em}No. of Prescribers by Specialty & 0.0267 (0.1868) \\
        \hspace{1em}No. of Pharmacies & 0.02 (0.1798) \\
        \hspace{1em}No. of Drug Products & 0.0231 (0.1501) \\
        \hspace{1em}Average Days Supply & 1.0728 (8.2557) \\
        \hline
        $^1$n(\%); mean(SD)
    \end{longtable}
\endgroup

The dosage variable was highly right skewed; while dosages in the full dataset ranged between 0 and 21,600 OMEs, around 80\% of refills had corresponding dosages under 500 OMEs. Dosage values were square-root transformed to correct for skewness, then normalized to lie between 0 and 1 to be used as the marks. Bandwidths for each mark-specific model were chosen to ensure that the data corresponding to refill events with marks falling within the bandwidth around each mark of interest contained sufficient information for model fitting. Ultimately, models were fit for marks 0.048, 0.068, 0.083, 0.118, and 0.144 (corresponding to dosages of 50, 100, 150, 300 and 450 OMEs), with bandwidths 0.01, 0.01, 0.01, 0.03, and 0.05, respectively.
Each fitted model adjusted for a comprehensive set of covariates, including demographic factors (age, sex, race/ethnicity, BMI), clinical characteristics (smoking, cancer diagnosis, comorbidities), surgical features (approach and inpatient versus outpatient setting), and pre-surgical prescription history. To obtain valid standard errors and confidence intervals, we used the proposed sandwich variance estimator. 

\subsection{Model Results and Interpretations}

\begin{figure}[ht]
    \centering
    \includegraphics[width=1\linewidth]{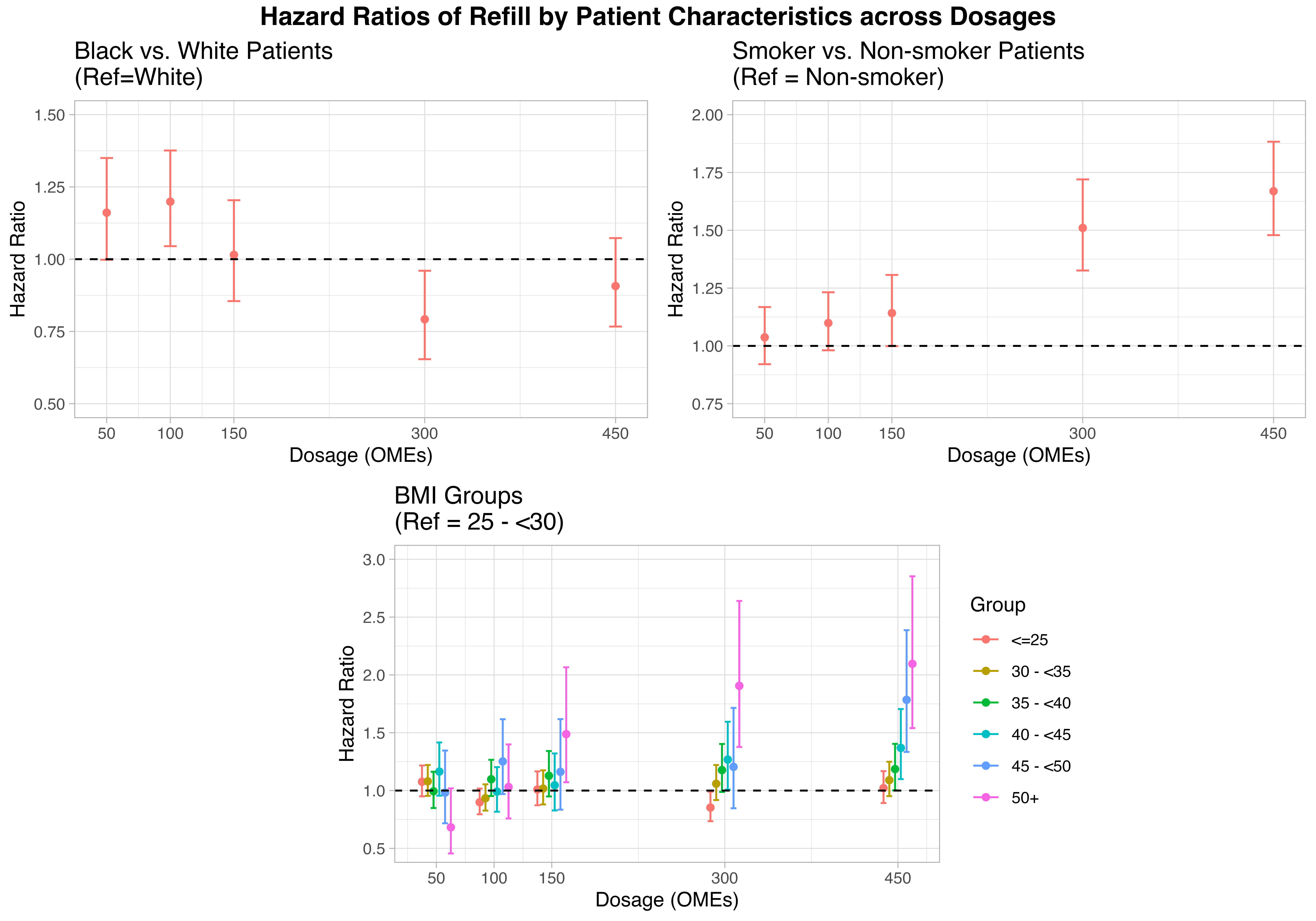}
    \caption{Adjusted hazard ratios of refill across opioid dosages for select patient characteristics. Plots of estimated mark-specific hazard ratios and corresponding 95\% CI error bars for race, smoking status, and BMI between 50 to 450 OMEs are shown here. The black dotted line in each plot represents the null hazard ratio (HR=1). Error bars crossing this line indicate non-significant results at the $\alpha=0.05$ level.}
    \label{fig:ptchar_res}
\end{figure}

\begin{figure}[ht]
    \centering
    \includegraphics[width=1\linewidth]{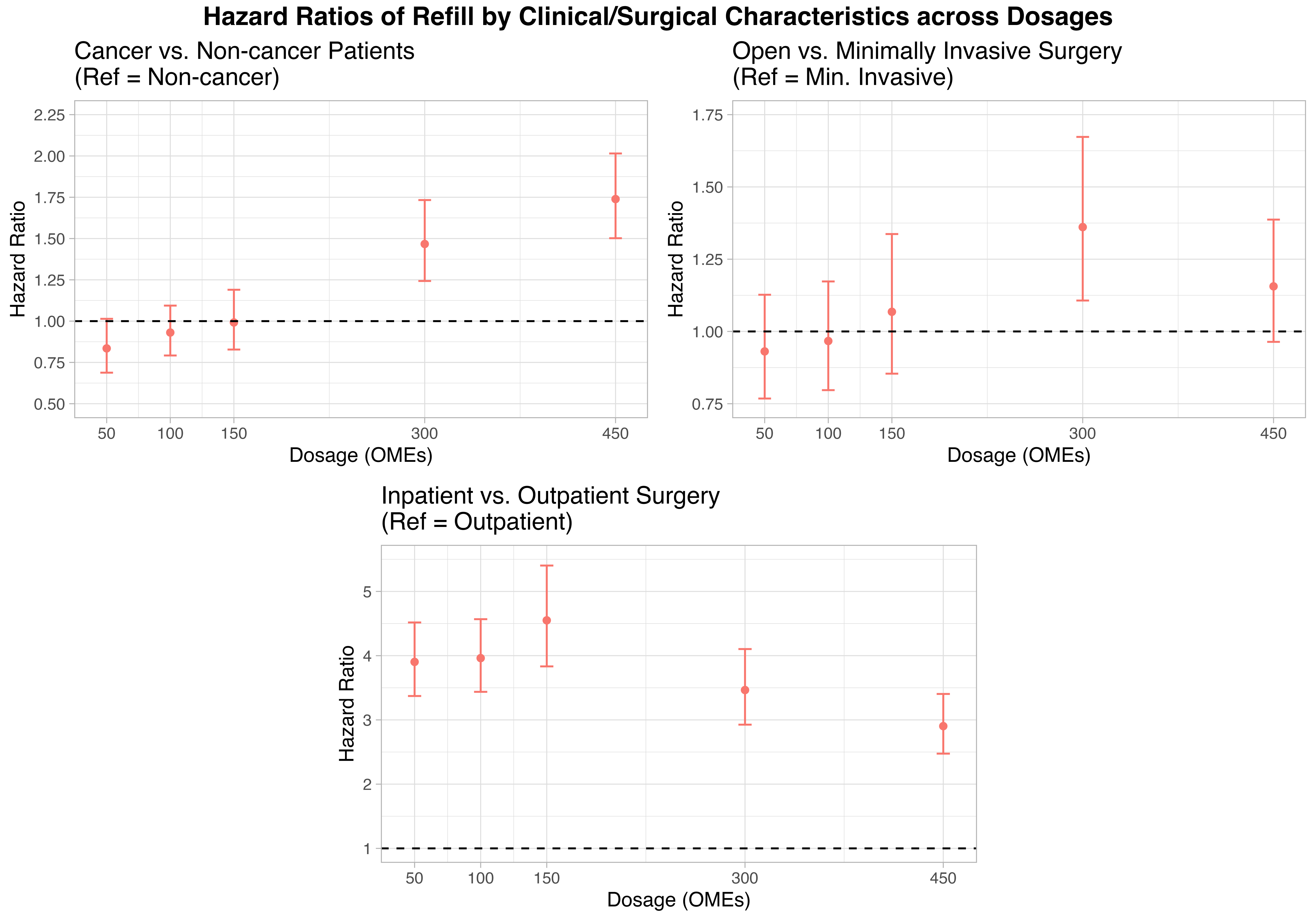}
        \caption{Adjusted hazard ratios of refill at opioid dosages between 50 to 450 OMEs for select clinical and surgical characteristics. Plots of estimated mark-specific hazard ratios and corresponding 95\% CI error bars for cancer status, surgical location, and surgical approach between 50 to 450 OMEs are shown here. The black dotted line in each plot represents the null hazard ratio (HR=1). Error bars crossing this line indicate non-significant results at the $\alpha=0.05$ level.}
    \label{fig:ptclin_res}
\end{figure}

Hazard ratios and 95\% CIs for the full covariate set from the mark-specific models (and from the non-mark-specific models for comparison) are reported in STable 1, and hazard ratios across dosages for key covariates are presented in Figures \ref{fig:ptchar_res} and \ref{fig:ptclin_res}. 
The results of our mark-specific models provide deeper insight into previously identified associations. Below, we present key findings for select demographic and clinical variables from our mark-specific model, while comparing them to the non-mark-specific model results and interpreting them in the context of prior literature. 

Among the patient demographic variables, our mark-specific model showed that Black patients had higher hazards of refill than White patients at lower dosages (adjusted HR (95\% CI): 1.16 (1.00, 1.35) at 50 OMEs, 1.20 (1.05, 1.38) at 100 OMEs, and 1.02 (0.86, 1.20) at 150 OMEs), but at higher dosages, Black patients had lower hazards of refill than White patients (HR (95\% CI): 0.79 (0.65, 0.96) at 300 OMEs and 0.91 (0.77, 1.07) at 450 OMEs), adjusting for all other covariates (STable 1, Figure \ref{fig:ptchar_res}). 
Considering the existing literature, a number of previous studies of postoperative pain after various procedures have identified more severe postoperative pain in Black patients than White patients \citep{thurston_systematic_2023}. However, Black patients have also been found be less likely to be prescribed opioid analgesics \citep{thurston_systematic_2023,morden_racial_2021}.
The non-mark-specific model we fit found that Black patients had 1.11 times the adjusted hazards of refill (Adjusted HR: 1.11; 95\% CI: 1.02, 1.20) ignoring dosage, compared to White patients (STable 1).
The mark-specific model results may provide further insight into the relationship between race and postoperative pain management. 
Higher hazards of refill for Black patients at lower dosages but lower hazards of refill at higher dosages may provide support for previous studies' findings that Black patients' pain is managed differently than White patients; perhaps Black patients experience higher levels of pain, but are not regularly prescribed higher dosages of opioid prescriptions after surgery. This could explain why they have lower hazards of high-dosage refills, but have higher hazards of low-dosage refills, as they may experience more frequent consumption of low-dosage prescriptions. Findings from \citet{morden_racial_2021} support this hypothesis; they found that Black patients not only had a lower rate of long-term opioid receipt, but Black patients had a lower annual mean dosage of opioid prescriptions than White patients within 91\% of the 310 health systems included in their study, which suggests different management of pain for Black patients relative to White patients.

From the mark-specific model, we also found that patients with higher BMI (namely those with BMI $\geq$ 40) had higher adjusted hazards of higher dosage refill events compared to average BMI patients (BMI=25-30), but their adjusted hazards of lower dosage refill events were not significantly different from average BMI patients (STable 1, Figure \ref{fig:ptchar_res}). Previous studies have identified links between obesity and the consumption of greater quantities of opioids in the postsurgical period \citep{howard_association_2019}, which aligns with the non-mark-specific model results that patients with BMI between 45-50 have significantly higher hazards of refill at any dosage level than patients with BMI between 25-30. The results from our mark-specific model could further suggest that patients with higher BMI may not only require larger opioid dosages, but also more frequent refills after surgery.

Smoking status is also known to be associated with postoperative opioid usage. \citet{wang_cigarette_2023} found that current smokers had higher levels of postoperative opioid consumption as well as pain scores than never smokers. Similarly, \citet{howard_association_2019} found that tobacco users used significantly more opioids than non-smokers in the postoperative period. \citet{montbriand_smoking_2018} found that smokers had significantly higher postoperative pain intensity and slower decline in daily opioid usage at 3 months after surgery than nonsmokers.
The non-mark-specific model results were consistent with greater opioid consumption among smokers identified the prior literature, with smokers having higher adjusted hazards of refill than non-smokers (HR (95\% CI): 1.17 (1.10, 1.25)). 
Our mark-specific model providers further evidence of a link between refill frequency and opioid dosage among smokers; while we found that smokers had higher hazards of refill than non-smokers at all dosage levels that were chosen for modeling, these differences were especially pronounced at higher dosage levels (adjusted HR (95\% CI): 1.51 (1.33, 1.72) at 300 OMEs and 1.67 (1.48, 1.88) at 450 OMEs) (STable 1, Figure \ref{fig:ptchar_res}). The upward trend in hazard ratios for refill as dosage level increases suggests that smoking is strongly indicative of higher dosages for prescription opioid refills.

Among the clinical and surgery-related variables, mark-specific hazards of refill at dosages of 50, 100 and 150 OMEs for cancer patients were not significantly different from those of non-cancer patients, but cancer patients had significantly higher hazards of refill at 300 and 450 OMEs than non cancer-patients (adjusting for other covariates) (STable 1, Figure \ref{fig:ptclin_res}). 
Meanwhile, the non-mark-specific model did not detect significantly different hazards of refill for cancer vs. non-cancer patients, thus failing to capture the significant results that the mark-specific model captured for refills of higher dosage levels (STable 1). 
Previous research has found that the risk of developing new persistent opioid usage is higher among cancer patients undergoing curative-intent surgery (10\% risk) than non-cancer patients undergoing surgical procedures (6-8\% risk) \citep{lee_new_2017,brummett_new_2017,soneji_risks_2016, clarke_rates_2014}. In \citet{salz_trends_2019}, researchers further found that cancer survivors who were chronic opioid users took higher dosages of opioids than non-cancer patients with chronic opioid usage in the first 3 to 5 years.
In the context of these previous findings, our mark-specific model results may suggest that cancer patients experience more frequent refills at higher dosage levels to manage their pain after surgery.

Surgical approach is another factor of interest in the study of postoperative opioid usage \citep{soneji_risks_2016, howard_association_2019}. In \citet{soneji_risks_2016}, a higher proportion of patients who received open surgeries had long-term ongoing opioid prescriptions than patients who received minimally invasive procedures. Furthermore, \citet{howard_association_2019} found that patients having open procedures reported higher consumption of opioids after surgery.
Our mark-specific models found similar results between surgical approach and postoperative opioid usage, and provided some further insight into this relationship. We found that patients who received an open surgery had (non-significantly) lower hazards of low-dosage refills, but significantly higher hazards of high-dosage refills than patients who had minimally invasive surgeries, especially at a dosage level of 300 OMEs (adjusted HR (95\% CI): 1.36 (1.11, 1.67)) (STable 1, Figure \ref{fig:ptclin_res}). This suggests that the association between open surgery and more frequent prescription refills is perhaps most pronounced at higher dosage levels. 

Finally, surgical location (inpatient versus outpatient procedure) was also associated with different hazards of postoperative opioid prescription refills in our analysis. Our mark-specific models revealed that patients who underwent an inpatient procedure had significantly higher adjusted hazards of refill than patients who underwent an outpatient procedure, across all of the studied dosage levels (STable 1, Figure \ref{fig:ptclin_res}). %Moreoever, the hazard ratios at each of the distinct dosage levels appeared to be similar, which may suggest that the relationship between surgical location and hazards of opioid prescription refill does not depend heavily on the opioid dosage. 
These results align with the findings from the studies conducted by \citet{varady_opioid_2021} and \citet{howard_association_2019}. \citet{varady_opioid_2021} found that while outpatient and inpatient total joint arthoplasty (TJA) patients receive prescriptions for similar dosages of opioids, patients undergoing inpatient procedures were significantly more likely to be on opioids 90 days after their procedure than patients who had received outpatient procedures. Similarly, \citet{howard_association_2019} found that inpatients typically consumed greater quantities of opioids than outpatients after surgery. 
\section{Discussion}

We have developed a method for modeling recurrent event hazards for events corresponding to continuous marks of interest. Many methods for modeling recurrent events have previously been developed, and Sun et al. (2009) developed an approach to model hazards for univariate survival due to a continuous mark of interest. However, to our knowledge, our method is the first to focus on modeling hazards for gap times in multivariate survival data (i.e., recurrent events) corresponding to continuous marks of interest associated with each event recurrence. Our estimation strategy addresses nonparametric modeling, marginal regression, and informative cluster size simultaneously. Furthermore, our derived sandwich estimator for variance estimation facilitates robust inference. 

 Many opportunities exist for refining and further testing the model. First,
 Eq. (\ref{cond mark pdf}) allows us to predict the dose level given the recurrent event time. We will explore this in detail in future work. 
 Also, while we focus on the marginal modeling approach in this paper, we anticipate extending our method to incorporate random effects to enable the direct modeling of within-subject correlations. 
 In this paper, we assume that subsequent gap times are not influenced by the marks of the preceding gaps, but this assumption may not hold in real applications. Thus, one extension of the current work could be to account for the informative nature of the mark on later gap times. 
 %Furthermore, the optimal choice of bandwidth for a mark of interest remains unclear. We envision that a K-fold cross-validation procedure can be adopted for choosing bandwidths for a given mark, and a more thorough investigation is needed. 
 Finally, in our estimating equations, we set the subject-specific weight $\omega_i$ to be the inverse of the number of uncensored gap times after the first gap time and achieve unbiased estimation. However, other  choices for this $\omega_i$ term may exist, warranting further exploration.

\newpage
\section*{Acknowledgments}
The work was supported by grants from NIH.

\section*{Disclosure Statement}
The authors declare  no competing interests.

\section*{Data Availability Statement}
The data used in this study are not publicly available due to privacy and confidentiality restrictions. Access to the datasets can  be obtained through a formal Data Use Agreement (DUA) process with the data custodians. 
\newpage

\bibliography{references}

\end{document}

\bibliographystyle{apalike}

\maketitle \author{}

\pagebreak

\section{Proofs of Theorem 1 and 2}
\subsection{Proof of Theorem 1 (Uniform Consistency)}

To show the uniform consistency of $\hat{\bm{\beta}}(v)$ with $\bm{\beta}(v)$, we need to show the assumptions of \textbf{Lemma A.1} from \cite{sun_proportional_2009} are met. This lemma extends \textbf{Theorem 5.7} of \cite{van_der_vaart_asymptotic_1998} on uniform consistency for random functions of $(\bm{\theta})$ to random functions of $(v,\bm{\theta})$ to accommodate the mark-specific setting.  

First, we define 
\[
    \begin{aligned}
    Q_n(v,\bm{\theta}) =
    \frac{1}{n} \sum_{i=1}^{n} \bigg [ \omega_i
    \sum_{j=1}^{R_i}
    \bigg\{
    \int_{0}^{1}
    \int_{0}^{\tau}
     K_h(u-v)
     \bigg [
     \bm{\theta}^\mathrm{T}\bm{Z}_i - 
    \log \big [ 
    \frac{1}{n} \sum_{k=1}^{n} \omega_k \sum_{l=1}^{R_k} Y_{k(l)}(t) \exp (\bm{\theta}^\mathrm{T} \bm{Z}_k )
    \big ] \bigg]%\\
     \times N_{i(j)}(dt, du)
     \bigg \}
     \bigg ].
    \end{aligned}
\]
Note that we can re-express this to show the connection between $l(v,\bm{\theta})$ and $Q_n(v,\bm{\theta})$:
\[
    \begin{aligned}
    \frac{1}{n} l(v,\bm{\theta})& =
    \frac{1}{n} \sum_{i=1}^{n} \bigg [ \omega_i
    \sum_{j=1}^{R_i}
    \bigg\{
    \int_{0}^{1}
    \int_{0}^{\tau}
     K_h(u-v)
     \bigg [
     \bm{\theta}^\mathrm{T}\bm{Z}_i - 
    \log \big [ 
   n* \frac{1}{n} \sum_{k=1}^{n} \omega_k \sum_{l=1}^{R_k} Y_{k(l)}(t) \exp (\bm{\theta}^\mathrm{T} \bm{Z}_k )
    \big ] \bigg]%\\
    \times N_{i(j)}(dt, du)
     \bigg \}
     \bigg ] \\ 
   %   &= \frac{1}{n} \sum_{i=1}^{n} \bigg [ \omega_i
   %  \sum_{j=1}^{R_i}
   %  \bigg\{
   %  \int_{0}^{1}
   %  \int_{0}^{\tau}
   %   K_h(u-v)
   %   \bigg [
   %   \bm{\theta}^\mathrm{T}\bm{Z}_i - 
   %  \log n  - 
   % \log \big[ \frac{1}{n} \sum_{k=1}^{n} \omega_k \sum_{l=1}^{R_k} Y_{k(l)}(t) \exp (\bm{\theta}^\mathrm{T} \bm{Z}_k ) \big]
   %  \bigg ] %\\&
   %   \times N_{i(j)}(dt, du)
   %   \bigg \}
   %   \bigg ]  \\
   %   &= \frac{1}{n} \sum_{i=1}^{n} \bigg [ \omega_i
   %  \sum_{j=1}^{R_i}
   %  \bigg\{
   %  \int_{0}^{1}
   %  \int_{0}^{\tau}
   %   K_h(u-v)
   %   \bigg [
   %   \bm{\theta}^\mathrm{T}\bm{Z}_i - 
   % \log \big[ \frac{1}{n} \sum_{k=1}^{n} \omega_k \sum_{l=1}^{R_k} Y_{k(l)}(t) \exp (\bm{\theta}^\mathrm{T} \bm{Z}_k ) \big]
   %  \bigg ] %\\&
   %   \times N_{i(j)}(dt, du)
   %   \bigg \}
   %   \bigg ] \\
   %   & - \frac{1}{n} \bigg[ \sum_{i=1}^{n} \omega_i \sum_{j=1}^{R_i} \big[ \int_{0}^{1} \int_{0}^{\tau} K_h(u-v) \log(n) \times N_{i(j)}(dt, du) \big] \bigg] \\
     &= Q_n(v,\bm{\theta}) - \frac{1}{n} \bigg[ \sum_{i=1}^{n} \omega_i \sum_{j=1}^{R_i} \big[ \int_{0}^{1} \int_{0}^{\tau} K_h(u-v) \log(n) \times N_{i(j)}(dt, du) \big] \bigg].
    \end{aligned}
\]
Then 
\[
    Q_n(v,\bm{\theta})=\frac{1}{n} l(v,\bm{\theta}) + \frac{1}{n} \bigg[ \sum_{i=1}^{n} \omega_i \sum_{j=1}^{R_i} \big[ \int_{0}^{1} \int_{0}^{\tau} K_h(u-v) \log(n) \times N_{i(j)}(dt, du) \big] \bigg].
\]

%\textbf{\textit{Step 2: Find the limit that $Q_n$ converges to ($Q$):}}

Next, let 
\[
        \eta_n(u,\bm{\theta}) = \frac{1}{n} \sum_{i=1}^{n} \bigg[ \omega_i \sum_{j=1}^{R_i}   \bigg\{ \int_{0}^{u}\int_{0}^{\tau} \big[ \bm{\theta}^\text{T}\bm{Z}_{i} - \log(S^{(0)}(t,\bm{\theta}))\big] N_{i(j)}(dt,du)\bigg\} \bigg]
\]
and
\[
        \xi_n(u,\bm{\theta}) = \frac{1}{n} \sum_{i=1}^{n} \bigg[ \omega_i \sum_{j=1}^{R_i}   \bigg\{ \int_{0}^{u}\int_{0}^{\tau} \big[ \bm{\theta}^\text{T}\bm{Z}_{i} - \log(s^{(0)}(t,\bm{\theta}))\big] N_{i(j)}(dt,du)\bigg\} \bigg].
\]

Regularity condition \textbf{C2} 
   %     $\sup_{t,\bm{\theta}} ||S^{(q)}(t,\bm{\theta}) - s^{(q)}(t,\bm{\theta})|| = O_p(n^{-1/2}) $. 
        implies  $S^{(0)}(t,\bm{\theta}) = s^{(0)}(t,\bm{\theta}) + O_p(n^{-1/2})$.
    Also,  $\eta_n(u,\bm{\theta}) = \xi_n(u,\bm{\theta}) + O_p(n^{-1/2})$ by Lipschitz continuity. Moreover, we have that $Q_n(v,\bm{\theta}) = \int_{0}^{1} K_h(u-v) \eta_n(du, \bm{\theta})$. 

Therefore,
\[
    \begin{aligned}
        Q_n(v,\bm{\theta})& = \int_{0}^{1}K_h(u-v)\eta_n(du,\bm{\theta}) \\
        &= \int_{0}^{1}K_h(u-v)\xi_n(du,\bm{\theta}) + O_p(n^{-1/2}h^{-1}) \text{ uniformly in } (v,\bm{\theta}) \in [0,1] \times [-B,B]^p \text{ for } B>0.
    \end{aligned}
\]

Also, applying the Glivenko-Cantelli and Donsker theorems,% $\xi_n$ is a $\mathbb{P}_n$-like object (with corresponding $\xi$ as $\mathbb{P}$, such that:

\[
    \begin{aligned}
        \xi_n(u,\bm{\theta}) = \xi(u, \bm{\theta}) + O_p(n^{-1/2}) \text{ uniformly in } (v,\bm{\theta}) \in [0,1] \times [-B,B]^p
    \end{aligned}
\]

where
\[
\begin{aligned}
        \xi(u, \bm{\theta}) = \EX[\omega_i
    \sum_{j=1}^{R_i}
    \int_{0}^{u}
    \int_{0}^{\tau}
     \big [
     \bm{\theta}^\mathrm{T}\bm{Z}_i - 
    \log (s^{(0)}(t, \bm{\theta})) \big] %\\
     \times N_{i(j)}(dt, du)
     ].
\end{aligned}
\]

We can rewrite this with conditional expectations:
\[
\begin{aligned}
        \xi(u, \bm{\theta}) &= \EX[\omega_i
    \sum_{j=1}^{R_i}
    \int_{0}^{u}
    \int_{0}^{\tau}
     \big [
     \bm{\theta}^\mathrm{T}\bm{Z}_i - 
    \log (s^{(0)}(t, \bm{\theta})) \big] %\\
     \times N_{i(j)}(dt, du)
     ] \\
     &= \EX \bigg[\EX\big[\omega_i
    \sum_{j=1}^{R_i}
    \int_{0}^{u}
    \int_{0}^{\tau}
     \big [
     \bm{\theta}^\mathrm{T}\bm{Z}_i - 
    \log (s^{(0)}(t, \bm{\theta})) \big] %\\
     \times N_{i(j)}(dt, du) | \bm{Z}_i,Y_i, \omega_i]\bigg] \\
     &=\EX \bigg[  \omega_i
    \sum_{j=1}^{R_i}
    \int_{0}^{u}
    \int_{0}^{\tau}
     \big [
     \bm{\theta}^\mathrm{T}\bm{Z}_i - 
    \log (s^{(0)}(t, \bm{\theta})) \big] %\\
    \lambda_0(t,u)\exp(\bm{\beta}(u)^\text{T} \bm{Z}_i) Y_{i(j)}(t) dt \space du \bigg].
\end{aligned}
\]

Therefore, 
\[
    \begin{aligned}
        Q_n(v,\bm{\theta})&= 
        \int_{0}^{1}K_h(u-v)\xi_n(du,\bm{\theta}) + O_p(n^{-1/2}h^{-1}) \\
        &=  \int_{0}^{1}K_h(u-v) \big[ \xi(du,\bm{\theta}) +O_p(n^{-1/2})\big] + O_p(n^{-1/2}h^{-1}) \\
        &= \int_{0}^{1}K_h(u-v)  \xi(du,\bm{\theta}) + O_p(n^{-1/2}h^{-1}). 
    \end{aligned}
\]

Letting $g(u)=\EX \bigg[  \omega_i
    \sum_{j=1}^{S_i}
    \int_{0}^{\tau}
     \big [
     \bm{\theta}^\mathrm{T}\bm{Z}_i - 
    \log (s^{(0)}(t, \bm{\theta})) \big] %\\
    \lambda_0(t,u)\exp(\bm{\beta}(u)^\text{T} \bm{Z}_i) Y_{i(j)}(t) dt  \bigg]$,
    we then use a u-substitution with $u'=\frac{u-v}{h}$ and a Taylor expansion of 
    $g(u)$
    %$Q(u,\bm{\theta})$ 
    about $v$ to get rid of the kernel integral in this expression:
\begin{equation*}
    \begin{aligned}
        Q_n(v,\bm{\theta})
        &= \int_0^1 h^{-1}hK(u')g(hu'+v)du'+O_p(n^{-1/2}h^{-1}) \\
        &= \int_0^1 K(u') \bigg[ g(v)-hu' \frac{d g(u)}{du}\bigg|_{u=v} + O(h^2) \bigg]du' + O_p(n^{-1/2}h^{-1})\\
        &= \int_0^1 K(u')g(v)du'- \int_0^1 K(u')u'g(v)du'+O(h^2)+O_p(n^{-1/2}h^{-1})\\
        &= g(v)+O_p(n^{-1/2}h^{-1})\\
        &=Q(v,\bm{\theta})+O_p(n^{-1/2}h^{-1}) \text{ uniformly in } (v,\bm{\theta}) \in [0,1] \times [-B,B]
    \end{aligned}
\end{equation*}
where
\[
     Q(v,\bm{\theta})=\EX \bigg[  \omega_i
    \sum_{j=1}^{R_i}
    \int_{0}^{\tau}
     \big [
     \bm{\theta}^\mathrm{T}\bm{Z}_i - 
    \log (s^{(0)}(t, \bm{\theta})) \big] %\\
    \lambda_0(t,v)\exp(\bm{\beta}(v)^\text{T} \bm{Z}_i) Y_{i(j)}(t) dt  \bigg].
\]

Furthermore, we can show that $s^{(q)}(t, \bm{\theta})$ and $Q(v,\bm{\theta})$, which uses averaged at-risk processes (i.e., counting and at-risk processes averaged across episodes within a subject), are equivalent to versions of these expressions that use subjects' first episode processes. Recall that within a subject $i$, the bivariate process $(T_{i(j)},V_{i(j)}), j=1, \ldots, R_i$ is identically distributed conditional on $(\bm{Z}_i, \bm{\gamma}_i)$, where $\bm{\gamma}_i$ represents latent subject-specific variables.

Letting $S^{*(q)}(t,\bm{\theta})=n^{-1} \sum_{i=1}^{n} [{Y_{i(1)} (t) \exp(\bm{\beta}(v)^\mathrm{T} \bm{Z}_i ) \bm{Z}_i}^{\otimes q} ], q=0,1,2$ and
$s^{*(q)}(t,\bm{\theta}) = \EX[s^{*(q)}(t,\bm{\theta})]$, 

\begin{equation*}
    \begin{aligned}
        s^{(q)}(t,\bm{\theta}) = \EX[S^{(q)}(t,\bm{\theta})] &=\EX[ \exp(\bm{\beta}(v)^\mathrm{T} \bm{Z}_i ) \bm{Z}_i^{\otimes q} \omega_i \sum_{j=1}^{R_i} \EX[{Y_{i(j)} (t)} | \bm{\gamma}_i,\bm{Z}_i] ] \\
        &= \EX[ \exp(\bm{\beta}(v)^\mathrm{T} \bm{Z}_i ) \bm{Z}_i^{\otimes q} \omega_i * R_i  \EX[{Y_{i(1)} (t)} | \gamma_i,\bm{Z}_i, R_i] ] \\
        &= \EX[S^{*(q)}(t,\bm{\theta})] \\
        &= s^{*(q)}(t,\bm{\theta}).
    \end{aligned}
\end{equation*}

Next, let 
$Q^*(v,\bm{\theta}) = \EX \big[  
    \int_{0}^{\tau}
     \big [
     \bm{\theta}^\mathrm{T}\bm{Z}_i - 
    \log (s^{*(0)}(t, \bm{\theta})) \big] %\\
    \lambda_0(t,v)\exp(\bm{\beta}(v)^\text{T} \bm{Z}_i) Y_{i(1)}(t) dt  \big]$.
Since $s^{*(q)}(t,\bm{\theta}) = s^{(q)}(t,\bm{\theta})$, 
\begin{equation*}
    \begin{aligned}
        Q(v,\bm{\theta}) &= \EX \big[  \omega_i \sum_{j=1}^{R_i}
    \int_{0}^{\tau}
     \big [
     \bm{\theta}^\mathrm{T}\bm{Z}_i - 
    \log (s^{*(0)}(t, \bm{\theta})) \big] %\\
    \lambda_0(t,v)\exp(\bm{\beta}(v)^\text{T} \bm{Z}_i) Y_{i(j)}(t) dt  \big] \\
    &= \EX \big[  \omega_i * R_i
    \int_{0}^{\tau}
     \big [
     \bm{\theta}^\mathrm{T}\bm{Z}_i - 
    \log (s^{*(0)}(t, \bm{\theta})) \big] %\\
    \lambda_0(t,v)\exp(\bm{\beta}(v)^\text{T} \bm{Z}_i)\EX \big[ Y_{i(1)}(t)  | \bm{\gamma}_i, \bm{Z}_i, R_i  \big] dt \big] \\
    &= \EX \big[ 
    \int_{0}^{\tau}
     \big [
     \bm{\theta}^\mathrm{T}\bm{Z}_i - 
    \log (s^{*(0)}(t, \bm{\theta})) \big] %\\
    \lambda_0(t,v)\exp(\bm{\beta}(v)^\text{T} \bm{Z}_i) Y_{i(1)}(t)   dt \big] \\
    &= Q^*(v,\bm{\theta})
    \end{aligned}
\end{equation*}

%\textit{\textbf{Step 3: Show $\bm{\beta}(v)$ maximizes $Q(v,\bm{\theta})$:}}
We next show $\bm{\beta}(v)$ maximizes $Q(v,\bm{\theta})$. To do so, we can show that $\bm{\beta}(v)$ is the well-separated point of maximum of $Q^*(v,\bm{\theta})$. Note that

\begin{equation*}
    \begin{aligned}
        \frac{\partial Q^*(v,\bm{\theta})}{\partial \bm{\theta}} &= \EX \bigg[ 
    \int_{0}^{\tau}
     \bigg [
     \bm{Z}_i - 
     \frac{s^{*(1)}(t, \bm{\theta})}{s^{*0)}(t, \bm{\theta})}
    \bigg] 
    \lambda_0(t,v)\exp(\bm{\beta}(v)^\text{T} \bm{Z}_i) Y_{i(1)}(t) dt \bigg] 
    \end{aligned}
\end{equation*}

and 
\begin{equation*}
    \begin{aligned}
        \frac{\partial^2 Q^*(v,\bm{\theta})}{\partial \bm{\theta}^2} = -\EX \bigg[  
    \int_{0}^{\tau}
     \bigg [
      \frac{s^{*(2)}(t, \bm{\theta})}{s^{*(0)}(t, \bm{\theta})} - 
      \bigg\{
     \frac{s^{*(1)}(t, \bm{\theta})}{s^{*(0)}(t, \bm{\theta})} \bigg\} ^{\otimes2}
    \bigg]
    \times
    \lambda_0(t,v)\exp(\bm{\beta}(v)^\text{T} \bm{Z}_i) Y_{i(1)}(t) dt \bigg],
    \end{aligned}
\end{equation*}
which is negative definite.
Now we can show that $\evalat {\frac{\partial Q^*(v,\bm{\theta})}{\partial \bm{\theta}}} {\bm{\theta}=\bm{\beta}(v)} = 0$:
\begin{equation*}
    \begin{aligned}
        \evalat {\frac{\partial Q^*(v,\bm{\theta})}{\partial \bm{\theta}}} {\bm{\theta}=\bm{\beta}(v)} &= \EX \bigg[ 
    \int_{0}^{\tau}
     \bigg [
     \bm{Z}_i - 
     \frac{s^{*(1)}(t, \bm{\beta}(v))}{s^{*(0)}(t, \bm{\beta}(v))}
    \bigg] 
    \lambda_0(t,v)\exp(\bm{\beta}(v)^\text{T} \bm{Z}_i) Y_{i(1)}(t) dt \bigg] \\
    &= \EX \bigg[ 
    \int_{0}^{\tau}
     %\bigg \{
     \bm{Z}_i \times \lambda_0(t,v)\exp(\bm{\beta}(v)^\text{T} \bm{Z}_i) Y_{i(1)}(t) dt %\bigg\} 
     \bigg] \\
     &\qquad - \EX \bigg[ 
    \int_{0}^{\tau}
     \bigg \{
     \frac{s^{*(1)}(t, \bm{\beta}(v))}{s^{*(0)}(t, \bm{\beta}(v))}
    \bigg\} \times \lambda_0(t,v)\exp(\bm{\beta}(v)^\text{T} \bm{Z}_i) Y_{i(1)}(t) dt\bigg] 
    \end{aligned}
\end{equation*}

Some algebra yields that
\begin{equation*}
    \begin{aligned}
        \EX \bigg[
    \int_{0}^{\tau}
     \bigg \{
     \frac{s^{*(1)}(t, \bm{\beta}(v))}{s^{*(0)}(t, \bm{\beta}(v))}
    \bigg\}  \lambda_0(t,v)\exp(\bm{\beta}(v)^\text{T} \bm{Z}_i) Y_{i(1)}(t) dt\bigg]  
    = \EX \bigg[
    \int_{0}^{\tau}
     \bm{Z}_i \lambda_0(t,v)\exp(\bm{\beta}(v)^\text{T} \bm{Z}_i) Y_{i(1)}(t) dt\bigg]
    \end{aligned}
\end{equation*}
which leads to $$\evalat {\frac{\partial Q^*(v,\bm{\theta})}{\partial \bm{\theta}}} {\bm{\theta}=\bm{\beta}(v)} =0. $$
Since the second derivative $\frac{\partial^2 Q(v,\bm{\theta})}{\partial \bm{\theta}^2}$ is negative definite, we have that $\bm{\beta}(v)$ maximizes $Q^{*}(v,\bm{\theta})$. Finally, since $Q(v,\bm{\theta}) = Q^*(v,\bm{\theta})$, $\bm{\beta}(v)$ maximizes $Q(v,\bm{\theta})$ as well.

%\textbf{\textit{Step 4: Put it all together}}
Since we showed earlier that $Q_n(v,\bm{\theta}) =Q(v,\bm{\theta})+o_p(n^{-1/2}h^{-1})$
uniformly in $(v,\bm{\theta}) \in [a,b] \times [-B,B]^p$, then we know $Q_n(v,\bm{\theta}) \xrightarrow{p}Q(v,\bm{\theta})$ uniformly in $(v,\bm{\theta}) \in [a,b] \times [-B,B]^p$. Similarly, we can show that $\frac{\partial Q_n(v,\bm{\theta})}{\partial \bm{\theta}} \xrightarrow{p}\frac{\partial Q(v,\bm{\theta})}{\partial \bm{\theta}}$ and $\frac{\partial^2 Q_n(v,\bm{\theta})}{\partial \bm{\theta}^2} \xrightarrow{p}\frac{\partial^2 Q(v,\bm{\theta})}{\partial \bm{\theta}^2}$, uniformly in $(v,\bm{\theta}) \in [a,b] \times [-B,B]^p$.
Additionally, each component of $\bm{\beta}(v)$ is bounded between $(-\tilde{B}, \tilde{B})$ given $\bm{\beta}(v)$ is bounded. 
Then we have that, $\forall \alpha > 0, \exists n_0 \text{ s.t. } \Pr (\hat{\bm{\beta}}(v) \in [-B, B]^p, a \leq v \leq b) > 1-\alpha \text{ for } n \geq n_0$.  
Therefore, by the checking of the conditions of \cite{sun_proportional_2009}'s \textbf{Lemma A.1} and the fact that $Q_n(v, \hat{\bm{\beta}}(v)) \geq Q_n(v, \bm{\beta}(v)),$ we have that
\[
    \begin{aligned}
        \forall \varepsilon>0 \text{, } &\Pr (\sup_{a \leq v \leq b} || \hat{\bm{\beta}}(v) - \bm{\beta}(v)|| > \varepsilon ) \\
        &= \alpha +  \Pr (\sup_{a \leq v \leq b} || \hat{\bm{\beta}}(v) - \bm{\beta}(v)|| > \varepsilon ,  \hat{\bm{\beta}}(v) \in [-B, B]^p, a \leq v \leq b) \\
       & \rightarrow \alpha \text { as } n \rightarrow \infty.
    \end{aligned}
\]
Finally, because  $\alpha$ is arbitrary, we have $$\Pr (\sup_{a \leq v \leq b} ||\hat{\bm{\beta}}(v) - \bm{\beta}(v)|| > \varepsilon) \rightarrow 0.  $$ 
\qedsymbol

\subsection{ Proof of Theorem 2 (Asymptotic Normality)}

% The outline of the proof of asymptotic normality of $\hat{\bm{\beta}}(v)$ is as follows:

% \begin{enumerate}
%     \item Express $\bm{U}(v, \hat{\bm{\beta}}(v))$ with $M_{i(j)}(dt,du)$ instead of $N_{i(j)}(dt,du)$ using a second-order Taylor expansion of $\lambda(t,v|z)$.
%     \item Replace $S^{(j)}$ with $s^{(j)}$. This lets us have iid clusters in the summation across subjects.
%     \item Apply the Lyapunov central limit theorem (assuming iid clusters for subjects) to get asymptotic distribution for the pseudo-score function
%     \item Find the sandwich variance of $\hat{\bm{\beta}}(v)$
% \end{enumerate}

% \noindent \textbf{The proof:}

% \textbf{\textit{Step 1: Re-express the pseudo-score function with $M_{i(j)}$}}.
We begin by noting that the pseudo-score equation can be expressed as follows:
\[
\begin{aligned}
    n^{-1/2} \bm{U}(v,\bm{\beta}(v)) &= n^{-1/2} \sum_{i=1}^{n}
    \Bigg[ \omega_i \sum_{j=1}^{R_i} \int_{0}^{1} \int_{0}^{\tau}
    K_h(u-v) \bigg[ \bm{Z}_i - \frac{S^{(1)}(t,\bm{\beta}(v))}{S^{(0)}(t,\bm{\beta}(v))}
    \bigg]  N_{i(j)}(dt,du)
    \Bigg] \\
    &= n^{-1/2} \sum_{i=1}^{n}
    \Bigg[ \omega_i \sum_{j=1}^{R_i} \int_{0}^{1} \int_{0}^{\tau}
    K_h(u-v) \bigg[ \bm{Z}_i - \frac{S^{(1)}(t,\bm{\beta}(v))}{S^{(0)}(t,\bm{\beta}(v))}
    \bigg]  
    \big[N_{i(j)}(dt,du) - Y_{i(j)}(t)\lambda(t,v|\bm{Z}_i) \big] dt du
    \Bigg]. 
\end{aligned}
\]

Note that from the second-order Taylor expansion of $\lambda(t,v|z)$, 
% \[
% \begin{aligned}
%     \lambda(t,u|\bm{Z}_i) &= \lambda(t,v|\bm{Z}_i) + \lambda'(t,v|\bm{Z}_i)(u-v)+\lambda''(t,v|\bm{Z}_i)\frac{(u-v)^2}{2!} \\
%     &= \lambda(t,v|\bm{Z}_i) + \lambda'(t,v|\bm{Z}_i)(u-v)+ o(4h^2/4)\\
%     &= \lambda(t,v|\bm{Z}_i) + \lambda'(t,v|\bm{Z}_i)(u-v)+o(h^2)
% \end{aligned}
% \]
% Then  
\[\lambda(t,u|\bm{Z}_i) - \lambda(t,v|\bm{Z}_i) = \lambda'(t,v|\bm{Z}_i)(u-v) + o(h^2).\]

Then, introduce  the mean-zero process $M_{i(j)}(dt,du)=N_{i(j)}(dt,du) - Y_{i(j)}(t)\lambda(t,u|\bm{Z}_i)  dt du$, the pseudo-score function can be expressed as follows:
\[
    \begin{aligned}
    &n^{-1/2} \bm{U}(v,\bm{\beta}(v))\\
    &= n^{-1/2} \sum_{i=1}^{n}
    \Bigg[ \omega_i \sum_{j=1}^{R_i} \int_{0}^{1} \int_{0}^{\tau}
    K_h(u-v) \bigg[ \bm{Z}_i - \frac{S^{(1)}(t,\bm{\beta}(v))}{S^{(0)}(t,\bm{\beta}(v))}
    \bigg]  
    M_{i(j)}(dt,du) \Bigg] %\blue{\text{ (pt 1)}}
    \\
    &\qquad+ n^{-1/2} \sum_{i=1}^{n}
    \Bigg[ \omega_i \sum_{j=1}^{R_i} \int_{0}^{1} \int_{0}^{\tau}
    K_h(u-v) \bigg[ \bm{Z}_i - \frac{S^{(1)}(t,\bm{\beta}(v))}{S^{(0)}(t,\bm{\beta}(v))}
    \bigg]  
    \times Y_{i(j)}(t)\{\lambda'(t,v|\bm{Z}_i)(u-v)\}  dt du \Bigg] %\blue{\text{ (pt 2)}}
    \\
    &\qquad+ n^{-1/2} \sum_{i=1}^{n}
    \Bigg[ \omega_i \sum_{j=1}^{R_i} \int_{0}^{1} \int_{0}^{\tau}
    K_h(u-v) \bigg[ \bm{Z}_i - \frac{S^{(1)}(t,\bm{\beta}(v))}{S^{(0)}(t,\bm{\beta}(v))}
    \bigg]  
    \times Y_{i(j)}(t)\{o(h^2)\}  dt du \Bigg], %\blue{\text{ (pt 3)}}
    \end{aligned}
\]
which can then be further simplified 
as
%using a u-substitution (letting $u'=(u-v)/h$) and taking into account the symmetric nature of the kernel function:
% \[
%     \begin{aligned}
%     &n^{-1/2} \bm{U}(v,\bm{\beta}(v))\\
%     &= n^{-1/2} \sum_{i=1}^{n}
%     \Bigg[ \omega_i \sum_{j=1}^{R_i} \int_{0}^{1} \int_{0}^{\tau}
%     K_h(u-v) \bigg[ \bm{Z}_i - \frac{S^{(1)}(t,\bm{\beta}(v))}{S^{(0)}(t,\bm{\beta}(v))}
%     \bigg]  
%     M_{i(j)}(dt,du) \Bigg] \blue{\text{ (pt 1)}}\\
%     &\qquad+ n^{-1/2} \sum_{i=1}^{n}
%     \Bigg[ \omega_i \sum_{j=1}^{R_i} \int_{-v/h}^{(1-v)/h} \int_{0}^{\tau}
%     h\frac{K(u')}{h} \bigg[ \bm{Z}_i - \frac{S^{(1)}(t,\bm{\beta}(v))}{S^{(0)}(t,\bm{\beta}(v))}
%     \bigg]  
%     \times Y_{i(j)}(t)\{\lambda'(t,v|\bm{Z}_i)(hu')\}  dt du' \Bigg] \blue{\text{ (pt 2)}}\\
%     &\qquad+ n^{-1/2} \sum_{i=1}^{n}
%     \Bigg[ \omega_i \sum_{j=1}^{R_i} \int_{-v/h}^{(1-v)/h} \int_{0}^{\tau}
%     h\frac{K(u')}{h} \bigg[ \bm{Z}_i - \frac{S^{(1)}(t,\bm{\beta}(v))}{S^{(0)}(t,\bm{\beta}(v))}
%     \bigg]  
%     \times Y_{i(j)}(t)\{o(h^2)\}  dt du' \Bigg] \blue{\text{ (pt 3)}}
%     \end{aligned}
% \]
% 
% \begin{itemize}
%     \item \blue{(pt 2)} in (22) will equal zero due to the symmetric nature of the kernel function: $\int K(u')u' du' = 0$.
%     \item \blue{(pt 3)} is equal to an $O_p(n^{1/2}h^2)$ term - we have $n$ $o(h^2)$ terms as we are summing over $n$ subjects, and $n^{-1/2} * n=n^{1/2}$. Hence we ultimately have an $O_p(n^{1/2}h^2)$ term. 
% \end{itemize}
% 
% So we end up with 
\[
    \begin{aligned}
    n^{-1/2} \bm{U}(v,\bm{\beta}(v))
    &= n^{-1/2} \sum_{i=1}^{n}
    \Bigg[ \omega_i \sum_{j=1}^{R_i} \int_{0}^{1} \int_{0}^{\tau}
    K_h(u-v) \bigg[ \bm{Z}_i - \frac{S^{(1)}(t,\bm{\beta}(v))}{S^{(0)}(t,\bm{\beta}(v))}
    \bigg]  M_{i(j)}(dt,du) \Bigg]  + O_p(n^{1/2}h^2).
    \end{aligned}
\]

% \textbf{\textit{Step 2: Replace $S^{(j)}$ with $s^{(j)}$}}

Next,  regularity condition \textbf{(C2)}  implies that 
 $\sup_{t \in[0,\tau], \bm{\theta}\in[-B,B]^p} ||S^{(q)}(t,\bm{\theta}) - s^{(q)}(t,\bm{\theta})|| = o_p(n^{-1/2+\delta})$, for $0<\delta<1/2.$ 
% This is because $n^{1/2} \text{ converges to 0 slower than } n^{1/4}$, so $n^{-1/2}$ converges to 0 faster than $n^{-1/4}$. So $o_p(n^{-1/2+0}) \Rightarrow o_p(n^{-1/2+1/4})$. (Since $n^{-1/2} < n^{-1/4}$).
Additionally, regularity condition \textbf{(C4)} implies that 
 $n^{-1/2+\delta} h^{-1/2}=o(1)$ for $\delta=1/4$, as $n^{-1/2+1/4}h^{-1/2}=(nh^2)^{-1/4}$, and if $nh^2 \rightarrow \infty$ then $(nh^2)^{-1/4} \rightarrow 0$.
Thus, we have: 
\[
    h^{1/2}K_h(u-v)||S^{(j)}(t,\bm{\beta}(v)) - s^{(j)}(t,\bm{\beta}(v))|| \xrightarrow{p}0. 
\]
% \begin{itemize}
%     \item Details: $K_h$ has $1/h$, the $||S-s||$ term is $o_p(n^{-1/4}$, so $h^{1/2}*h^{-1}*o_p(n^{-1/4}h^{1/2})=o_p((nh^2)^{-1/4})=o_p(1)$ 
% \end{itemize}
Then, using Lemma 2 of \cite{gilbert_2-sample_2008} and the facts that $S^{(1)}$ and $S^{(0)}$ both converge and $S^{(0)}$ is bounded away from 0, we have: 
% \begin{itemize}
%     \item Note that since $S^{(1)}$ and $S^{(0)}$ both converge and $S^{(0)}$ is bounded away from 0, $\sup_{t \in [0,\tau], \bm{\theta} \in [-B,B]^p}||\frac{S^{(1)}}{S^{(0)}}- \frac{s^{(1)}}{s^{(0)}}|| \xrightarrow{p} 0$. 
% \end{itemize}
% \[
% \begin{aligned}
%     &h^{1/2}K_h(u-v)||\frac{S^{(1)}}{S^{(0)}}- \frac{s^{(1)}}{s^{(0)}}|| \xrightarrow{p}0 \\
%     & \Rightarrow h^{1/2}K_h(u-v)\bigg[\frac{S^{(1)}}{S^{(0)}} \bigg] = h^{1/2}K_h(u-v)\bigg[\frac{s^{(1)}}{s^{(0)}} \bigg] + o_p(1)
% \end{aligned}
% \]
% So, applying Lemma 2 to the pseudo-score function, we get:
\[
    \begin{aligned}
        &n^{-1/2}h^{1/2}\bm{U}(v,\bm{\beta}(v)) \\ & \qquad = 
        n^{-1/2}h^{1/2} \sum_{i=1}^{n} \Bigg[ \omega_i \sum_{j=1}^{R_i} \int_{0}^{1} \int_{0}^{\tau}
    K_h(u-v) \bigg[ \bm{Z}_i - \frac{S^{(1)}(t,\bm{\beta}(v))}{S^{(0)}(t,\bm{\beta}(v))}
    \bigg]  M_{i(j)}(dt,du) \Bigg] + O_p(n^{1/2}h^{5/2}) \\ & \qquad = 
        n^{-1/2}h^{1/2} \sum_{i=1}^{n} \Bigg[ \omega_i \sum_{j=1}^{R_i} \int_{0}^{1} \int_{0}^{\tau}
    K_h(u-v) \bigg[ \bm{Z}_i - \frac{s^{(1)}(t,\bm{\beta}(v))}{s^{(0)}(t,\bm{\beta}(v))}
    \bigg]  M_{i(j)}(dt,du) \Bigg] + O_p(n^{1/2}h^{5/2}) + o_p(1).
    \end{aligned}
\]

% Next, we need to replace the $v$ with $u$ in $\frac{S^{(1)}(t,\bm{\beta}(v))}{S^{(0)}(t,\bm{\beta}(v))}$. To do this, we can again use a Taylor Expansion and u-substitution.

% \pink{
% \begin{itemize}
%     \item Let $u'=(u-v)/h$ again...
%     \item Then the "leftover" terms from doing the Taylor Expansion fold into the existing $O_p(n^{1/2}h^{5/2})$ term so that term remains unchanged.
% \end{itemize}
% }

% Then we have:
% \[
%     \begin{aligned}
%         &n^{-1/2}h^{1/2}\bm{U}(v,\bm{\beta}(v)) \\ &\qquad = 
%         n^{-1/2}h^{1/2} \sum_{i=1}^{n} \Bigg[ \omega_i \sum_{j=1}^{R_i} \int_{0}^{1} \int_{0}^{\tau}
%     K_h(u-v) \bigg[ \bm{Z}_i - \frac{s^{(1)}(t,\bm{\beta}(v))}{s^{(0)}(t,\bm{\beta}(v))}
%     \bigg]  M_{i(j)}(dt,du) \Bigg] + O_p(n^{1/2}h^{5/2}) + o_p(1).
%     \end{aligned}
% \]
% 
%\textbf{\textit{Step 3: Central Limit Theorem for Pseudo-score}}

 We now apply the Lyapunov Central Limit Theorem to the pseudo-score equation, recognizing that the distribution of event clusters across subjects depends on the sample size \( n \), and that the bandwidth \( h \) also varies with \( n \). To verify the conditions of the Lyapunov CLT, we first observe that the data from each subject are  independent and identically distributed (i.i.d.) across subjects, satisfying the independence requirement.  We  further examine the third moment of the pseudo-score function and confirm that it remains finite as \( n \to \infty \), based on the contributions of the relevant terms.
% Note that:
% \begin{itemize}
%     \item We treat the clusters of events per subject as iid now rather than the events themselves
%     \item Our $h$ is dependent on $n$ (see regularity conditions). This means that the distribution of the clusters depends on sample size, so we need the Lyapunov CLT instead of the traditional CLT. \blue{(Relates to triangular array issues--for the traditional CLT to be used, we would need to know that no matter what the $n$ is, the individual $X_i, i=1,\ldots n$ have the same distribution).}
%     \begin{itemize}
%         \item Lyapunov conditions: $X_i$ are independent but not necessarily iid, $X_i, i=1,\ldots n, \ldots$ have finite mean $\mu_i$ and variance $\sigma^2_i$, and there's a finite 3rd moment.
%     \end{itemize}
% \end{itemize}
%\textit{Checking Lyapunov conditions:}
%$\rightarrow$ 3rd moment condition:
%To check the Lyapunov condition, we need to make sure $\EX [(n^{-1/2}h^{1/2}\bm{U}(v,\bm{\beta}(v))^3]$ is finite.
Indeed, we can show
\begin{equation*} 
    \begin{aligned}
    &\EX [  (n^{-1/2}h^{1/2}\bm{U}(v,\bm{\beta}(v))^3] \\ & \qquad =
    \EX \bigg[ \bigg\{ n^{-1/2}h^{1/2} \sum_{i=1}^{n} \Bigg[ \omega_i \sum_{j=1}^{R_i} \int_{0}^{1} \int_{0}^{\tau}
    K_h(u-v) \bigg[ \bm{Z}_i - \frac{s^{(1)}(t,\bm{\beta}(v))}{s^{(0)}(t,\bm{\beta}(v))}
    \bigg]  M_{i(j)}(dt,du) \Bigg] + O_p(n^{1/2}h^{5/2}) + o_p(1) 
    \bigg\}^3
    \bigg] \\  
    & = (nh^3)^{-1/2} + o_p(1).
    \end{aligned}
    \end{equation*}
      which will go to zero since $nh^3 \rightarrow \infty$ as $n \rightarrow \infty$ (see regularity conditions).
    Hence, the Lyapunov Central Limit Theorem implies that 
    $ n^{-1/2}h^{1/2}\bm{U}(v,\bm{\beta}(v))$ will converge weakly to a mean zero multivariate normal distribution.
To complete the proof, we derive the asymptotic variance of $n^{-1/2}h^{1/2}\bm{U}(v,\bm{\beta}(v))$. Note that:
\[
  Var(n^{-1/2}h^{1/2}\bm{U}(v,\bm{\beta}(v)))
  %=\frac{1}{n}h  n Var(\bm{U}_i(v,\bm{\beta}(v)))
  =hVar(\bm{U}_i(v,\bm{\beta}(v))) 
  =\EX[h\bm{U}_i(v,\bm{\beta}(v))\bm{U}_i(v,\bm{\beta}(v))^{\text{T}}]
\]
where \[
\bm{U}_i(v,\bm{\beta}(v))) = \omega_i \sum_{j=1}^{R_i} \int_{0}^{1} \int_{0}^{\tau}
    K_h(u-v) \bigg[ \bm{Z}_i - \frac{s^{(1)}(t,\bm{\beta}(v))}{s^{(0)}(t,\bm{\beta}(v))}
    \bigg]  M_{i(j)}(dt,du).\]
More explicitly,
\begin{equation} \label{pseudoscorevar}
    \begin{aligned} 
    & \EX[ h \bm{U}_i(v,\bm{\beta}(v))\bm{U}_i(v,\bm{\beta}(v))^{\text{T}}] \\ &  = 
         \EX \bigg[ h 
        \omega_i^2 
        \bigg\{ \sum_{j=1}^{R_i} \int_{0}^{1} \int_{0}^{\tau}
    K_h(u-v) \bigg[ \bm{Z}_i - \frac{s^{(1)}(t,\bm{\beta}(v))}{s^{(0)}(t,\bm{\beta}(v))}
    \bigg]  M_{i(j)}(dt,du) \bigg\} \\& \qquad \qquad
    \bigg\{ \sum_{j=1}^{R_i} \int_{0}^{1} \int_{0}^{\tau}
    K_h(u-v) \bigg[ \bm{Z}_i - \frac{s^{(1)}(t,\bm{\beta}(v))}{s^{(0)}(t,\bm{\beta}(v))}
    \bigg]  M_{i(j)}(dt,du) \bigg\}^{\text{T}}
        \bigg] \\
        &= \EX\bigg[ h \omega_i^2  \sum_{j=1}^{R_i}  \sum_{k=1}^{R_i} 
        \bigg\{\int_{0}^{1} \int_{0}^{\tau}
    K_h(u-v) \bigg[ \bm{Z}_i - \frac{s^{(1)}(t,\bm{\beta}(v))}{s^{(0)}(t,\bm{\beta}(v))}
    \bigg]  M_{i(j)}(dt,du) \bigg\} \\& \qquad \qquad
        \bigg\{\int_{0}^{1} \int_{0}^{\tau}
    K_h(u-v) \bigg[ \bm{Z}_i - \frac{s^{(1)}(t,\bm{\beta}(v))}{s^{(0)}(t,\bm{\beta}(v))}
    \bigg]  M_{i(k)}(dt,du) \bigg\}^{\text{T}}
        \bigg] 
        \\
        &=\EX\bigg[ \omega_i^2  \sum_{j=1}^{R_i}  \sum_{k=1}^{R_i} \bigg\{\int_{0}^{1} \int_{0}^{\tau} \int_{0}^{1} \int_{0}^{\tau} h
        K_h(u_1-v)  K_h(u_2-v) 
        \bigg[ \bm{Z}_i - \frac{s^{(1)}(t_1,\bm{\beta}(v))}{s^{(0)}(t_1,\bm{\beta}(v))}
    \bigg] 
    \bigg[ \bm{Z}_i - \frac{s^{(1)}(t_2,\bm{\beta}(v))}{s^{(0)}(t_2,\bm{\beta}(v))} \bigg]^\text{T}
    \\ & \qquad \qquad \qquad \times M_{i(j)}(dt_1, du_1) M_{i(k)}(dt_2, du_2)
        \bigg\}
         \bigg].
    \end{aligned}
\end{equation}

Now, to evaluate this expression, we consider two separate cases corresponding to the value of $R_i$. 
Recall that if $R_i=1$, censoring is possible and there is only one episode (censored or uncensored) per subject, so $N_{i(1)}(t,v)=I(X_{i(1)}\leq t, \delta_{i(1)} = 1, V_{i(1)} \leq v)$ and $Y_{i(1)}(t) = I(X_{i(1)}(t) \geq t)$. 
%Additionally, $\EX[{N_{i(1)}(dt,dv)|\bm{Z}_i,Y_{i(1)}(t)}] = Y_{i(1)}(t)\lambda(t,v|\bm{Z}_i)dtdv$. 
If $R_i>1$, there is more than one gap for subject $i$ and there is no censoring for any gap times for that subject. Then $N_{i(j)}(t,v)=I(T_{i(j)}\leq t, V_{i(j)} \leq v)$ and $Y_{i(j)}(t) = I(T_{i(j)} \geq t)$.

We break down (\ref{pseudoscorevar}) conditioning on $R_i$:
\begin{equation*} 
    \begin{aligned}
    &\EX[ h \bm{U}_i(v,\bm{\beta}(v))\bm{U}_i(v,\bm{\beta}(v))^{\text{T}}]\\
    &= \EX[ \EX[ h \bm{U}_i(v,\bm{\beta}(v))\bm{U}_i(v,\bm{\beta}(v))^{\text{T}} | R_i]]\\
         &= 
         \EX\Bigg[ \EX\bigg[
         \omega_i^2  \sum_{j=1}^{R_i}  \sum_{k=1}^{R_i} \bigg\{\int_{0}^{1} \int_{0}^{\tau} \int_{0}^{1} \int_{0}^{\tau} h
        K_h(u_1-v)  K_h(u_2-v) 
        \bigg[ \bm{Z}_i - \frac{s^{(1)}(t_1,\bm{\beta}(v))}{s^{(0)}(t_1,\bm{\beta}(v))}
    \bigg] 
    \bigg[ \bm{Z}_i - \frac{s^{(1)}(t_2,\bm{\beta}(v))}{s^{(0)}(t_2,\bm{\beta}(v))} \bigg]^\text{T}
    \\ &\qquad \qquad \qquad \times M_{i(j)}(dt_1, du_1) M_{i(k)} (dt_2,du_2)
        \bigg\} \bigg| R_i
         \bigg]\Bigg]
    \end{aligned}
\end{equation*}

First, we consider the case where $R_i=1$:
\begin{equation*}
    \begin{aligned}
    &\EX[ h \bm{U}_i(v,\bm{\beta}(v))\bm{U}_i(v,\bm{\beta}(v))^{\text{T}}|R_i=1] \\ &= 
        \EX\bigg[\bigg\{
         \int_{0}^{1} \int_{0}^{\tau} \int_{0}^{1} \int_{0}^{\tau} h
        K_h(u_1-v)  K_h(u_2-v) 
        \bigg[ \bm{Z}_i - \frac{s^{(1)}(t_1,\bm{\beta}(v))}{s^{(0)}(t_1,\bm{\beta}(v))}
    \bigg] 
    \bigg[ \bm{Z}_i - \frac{s^{(1)}(t_2,\bm{\beta}(v))}{s^{(0)}(t_2,\bm{\beta}(v))} \bigg]^\text{T}
    \\ &\qquad \qquad \qquad \qquad \times N_{i(1)}(dt_1, du_1) N_{i(1)} (dt_2,du_2) 
        \bigg\} \bigg| R_i=1 \bigg] %\blue{\text{(pt a)}} 
        \\
    &\qquad  - 
    \EX\bigg[\bigg\{
         \int_{0}^{1} \int_{0}^{\tau} \int_{0}^{1} \int_{0}^{\tau} h
        K_h(u_1-v)  K_h(u_2-v) 
        \bigg[ \bm{Z}_i - \frac{s^{(1)}(t_1,\bm{\beta}(v))}{s^{(0)}(t_1,\bm{\beta}(v))}
    \bigg] 
    \bigg[ \bm{Z}_i - \frac{s^{(1)}(t_2,\bm{\beta}(v))}{s^{(0)}(t_2,\bm{\beta}(v))} \bigg]^\text{T}
    \\ &\qquad \qquad \qquad \qquad \times N_{i(1)}(dt_1, du_1)Y_{i(1)}(t_2)\lambda(t_2,u_2|\bm{Z}_i)dt_2 du_2 
        \bigg\}  \bigg| R_i=1 \bigg] 
        %\blue{\text{(pt b)}}
        \\
    &\qquad -
    \EX\bigg[\bigg\{
         \int_{0}^{1} \int_{0}^{\tau} \int_{0}^{1} \int_{0}^{\tau} h
        K_h(u_1-v)  K_h(u_2-v) 
        \bigg[ \bm{Z}_i - \frac{s^{(1)}(t_1,\bm{\beta}(v))}{s^{(0)}(t_1,\bm{\beta}(v))}
    \bigg] 
    \bigg[ \bm{Z}_i - \frac{s^{(1)}(t_2,\bm{\beta}(v))}{s^{(0)}(t_2,\bm{\beta}(v))} \bigg]^\text{T}
    \\ &\qquad \qquad \qquad \qquad \times N_{i(1)}(dt_2,du_2)Y_{i(1)}(t_1)\lambda(t_1,u_1|\bm{Z}_i)dt_1 du_1  
        \bigg\}  \bigg| R_i=1 \bigg] %\blue{\text{(pt c)}} 
        \\
    &\qquad +
    \EX\bigg[\bigg\{
         \int_{0}^{1} \int_{0}^{\tau} \int_{0}^{1} \int_{0}^{\tau} h
        K_h(u_1-v)  K_h(u_2-v) 
        \bigg[ \bm{Z}_i - \frac{s^{(1)}(t_1,\bm{\beta}(v))}{s^{(0)}(t_1,\bm{\beta}(v))}
    \bigg] 
    \bigg[ \bm{Z}_i - \frac{s^{(1)}(t_2,\bm{\beta}(v))}{s^{(0)}(t_2,\bm{\beta}(v))} \bigg]^\text{T}
    \\ &\qquad \qquad \qquad \qquad \times Y_{i(1)}(t_1)\lambda(t_1,u_1|\bm{Z}_i) Y_{i(1)}(t_2)\lambda(t_2,u_2|\bm{Z}_i)dt_1du_1 dt_2du_2  
        \bigg\}  \bigg| R_i=1 \bigg]  %\blue{\text{(pt d)}}
    \end{aligned}
\end{equation*}

%\ylcm{so the trick is only when $u_1=u_2$,  $h \rightarrow 0$ makes the cross-product  have a non zero limit; otherwise the  limit of the cross-product is zero.}

%where $g_{jk}(t_1, u_1, t_2, u_2|\bm{Z}_i) = \EX[M_{i(j)}(t_1,u_1) M_{i(k)}(t_2,u_2)|\bm{Z}_i] = Cov(M_{i(j)}(t_1,u_1), M_{i(k)}(t_2,u_2))$. Note that $\frac{\partial ^4 g_{jk}(t_1,u_1, t_2, u_2 |\bm{Z}_i)}{\partial t_1 u_1 t_2 u_2} dt_1 du_1 dt_2 du_2 = Cov(M_{i(j)}(dt_1,du_1), M_{i(k)}(dt_2,du_2))$.

%Then, we consider what happens as $h \rightarrow 0$. 

\noindent Noting that $N_{i(1)}(dt_1, du_1)N_{i(1)}(dt_2, du_2)$ only experiences jumps when $t_1=t_2, u_1=u_2$, and using the law of total expectation, this simplifies to the following:
\begin{equation*}
    \begin{aligned}
        &\EX[ h \bm{U}_i(v,\bm{\beta}(v))\bm{U}_i(v,\bm{\beta}(v))^{\text{T}}|R_i=1] \\ &= \EX\bigg[ 
        \bigg\{\int_{0}^{1} \int_{0}^{\tau} h
        K_h^2(u_1-v) 
        \bigg[ \bm{Z}_i - \frac{s^{(1)}(t_1,\bm{\beta}(v))}{s^{(0)}(t_1,\bm{\beta}(v))}
    \bigg] 
    \bigg[ \bm{Z}_i - \frac{s^{(1)}(t_1,\bm{\beta}(v))}{s^{(0)}(t_1,\bm{\beta}(v))} \bigg]^\text{T} \\
    &\qquad \qquad\qquad\qquad \times f(t_1,u_1|X_{i(1)} \geq t_1, \delta_{i(1)}=1, \bm{Z}_i, R_i = 1) dt_1 du_1 \bigg\} \bigg| R_i = 1 \bigg] \\
    & \qquad - \EX\bigg[ 
        \bigg\{\int_{0}^{1} \int_{0}^{\tau} \int_{0}^{1} \int_{0}^{\tau} h
        K_h(u_1-v)  K_h(u_2-v) 
        \bigg[ \bm{Z}_i - \frac{s^{(1)}(t_1,\bm{\beta}(v))}{s^{(0)}(t_1,\bm{\beta}(v))}
    \bigg] 
    \bigg[ \bm{Z}_i - \frac{s^{(1)}(t_2,\bm{\beta}(v))}{s^{(0)}(t_2,\bm{\beta}(v))} \bigg]^\text{T} \\ &\qquad \qquad\qquad\qquad \times Y_{i(1)}(t_1) f(t_1,u_1|X_{i(1)} \geq t_1, \delta_{i(1)}=1, \bm{Z}_i, R_i = 1) Y_{i(1)}(t_2) \lambda(t_2,u_2|\bm{Z}_i) dt_1 du_1 dt_2 du_2 \bigg\}  \bigg| R_i = 1  \bigg] \\ 
    & \qquad - \EX\bigg[ 
        \bigg\{\int_{0}^{1} \int_{0}^{\tau} \int_{0}^{1} \int_{0}^{\tau} h
        K_h(u_1-v)  K_h(u_2-v) 
        \bigg[ \bm{Z}_i - \frac{s^{(1)}(t_1,\bm{\beta}(v))}{s^{(0)}(t_1,\bm{\beta}(v))}
    \bigg] 
    \bigg[ \bm{Z}_i - \frac{s^{(1)}(t_2,\bm{\beta}(v))}{s^{(0)}(t_2,\bm{\beta}(v))} \bigg]^\text{T} \\ &\qquad \qquad\qquad\qquad \times Y_{i(1)}(t_1) \lambda(t_1,u_1|\bm{Z}_i) Y_{i(1)}(t_2) f(t_2,u_2|X_{i(1)} \geq t_2, \delta_{i(1)}=1, \bm{Z}_i, R_i = 1) dt_1 du_1 dt_2 du_2 \bigg\}  \bigg| R_i = 1  \bigg] \\
    & \qquad + \EX\bigg[\bigg\{
         \int_{0}^{1} \int_{0}^{\tau} \int_{0}^{1} \int_{0}^{\tau} h
        K_h(u_1-v)  K_h(u_2-v) 
        \bigg[ \bm{Z}_i - \frac{s^{(1)}(t_1,\bm{\beta}(v))}{s^{(0)}(t_1,\bm{\beta}(v))}
    \bigg] 
    \bigg[ \bm{Z}_i - \frac{s^{(1)}(t_2,\bm{\beta}(v))}{s^{(0)}(t_2,\bm{\beta}(v))} \bigg]^\text{T}
    \\ &\qquad \qquad \qquad \qquad \times Y_{i(1)}(t_1)\lambda(t_1,u_1|\bm{Z}_i) Y_{i(1)}(t_2)\lambda(t_2,u_2|\bm{Z}_i)dt_1du_1 dt_2du_2  
        \bigg\}  \bigg| R_i=1 \bigg] 
    \end{aligned}   
\end{equation*}

\noindent Note that the last three terms in the expression above converge to zero as $h \rightarrow 0$; this can be seen using u-substitution with $u_1'=\frac{u_1-v}{h}, u_2' = \frac{u_2-v}{h}$, and taking the limit as $h \rightarrow 0$. Then, letting $\nu_0 = \int K^2(u)du  \text{ (=3/5 for the Epanechnikov kernel})$, we have:
\begin{equation*}
    \begin{aligned}
        &\EX[ h \bm{U}_i(v,\bm{\beta}(v))\bm{U}_i(v,\bm{\beta}(v))^{\text{T}}|R_i=1] \\ & \qquad \rightarrow \nu_0 \EX\bigg[
         \bigg\{\int_{0}^{\tau} 
        \bigg[ \bm{Z}_i - \frac{s^{(1)}(t_1,\bm{\beta}(v))}{s^{(0)}(t_1,\bm{\beta}(v))}
    \bigg] 
    \bigg[ \bm{Z}_i - \frac{s^{(1)}(t_1,\bm{\beta}(v))}{s^{(0)}(t_1,\bm{\beta}(v))} \bigg]^\text{T} f(t_1,v|X_{i(1)} \geq t_1, \delta_{i(1)}=1, \bm{Z}_i, R_i = 1) dt_1  \bigg\} \bigg| R_i = 1 \bigg] \\
    & \qquad \quad (\equiv \bm{\Lambda}_1(v)).
    \end{aligned}
\end{equation*}

Next, we examine the case where $R_i=r$, for $r>1$, considering when $j=k$ and $j \neq k$ in the expression separately:
\begin{equation*}
    \begin{aligned}
        &\EX[ h \bm{U}_i(v,\bm{\beta}(v))\bm{U}_i(v,\bm{\beta}(v))^{\text{T}}|R_i=r] \\
        &  = \EX\bigg[
         r^{-2} \sum_{j=1}^{r} \bigg\{\int_{0}^{1} \int_{0}^{\tau} \int_{0}^{1} \int_{0}^{\tau} h
        K_h(u_1-v)  K_h(u_2-v) 
        \bigg[ \bm{Z}_i - \frac{s^{(1)}(t_1,\bm{\beta}(v))}{s^{(0)}(t_1,\bm{\beta}(v))}
    \bigg] 
    \bigg[ \bm{Z}_i - \frac{s^{(1)}(t_2,\bm{\beta}(v))}{s^{(0)}(t_2,\bm{\beta}(v))} \bigg]^\text{T}
    \\ &\qquad \qquad \qquad \qquad \qquad  \qquad  \qquad \times M_{i(j)}(dt_1, du_1) M_{i(j)} (dt_2,du_2) \bigg\} \bigg| R_i=r
        \bigg] %\text{\orange{(case 2.1)}}
        \\
        & \quad  + 
        \EX\bigg[
         r^{-2}  \sum_{j\neq k }^{r}  \sum_{k=1}^{r} \bigg\{\int_{0}^{1} \int_{0}^{\tau} \int_{0}^{1} \int_{0}^{\tau} h
        K_h(u_1-v)  K_h(u_2-v) 
        \bigg[ \bm{Z}_i - \frac{s^{(1)}(t_1,\bm{\beta}(v))}{s^{(0)}(t_1,\bm{\beta}(v))}
    \bigg] 
    \bigg[ \bm{Z}_i - \frac{s^{(1)}(t_2,\bm{\beta}(v))}{s^{(0)}(t_2,\bm{\beta}(v))} \bigg]^\text{T}
    \\ &\qquad \qquad \qquad \qquad \qquad  \qquad  \qquad \times M_{i(j)}(dt_1, du_1) M_{i(k)} (dt_2,du_2) \bigg\} \bigg| R_i=r
        \bigg] 
        %\text{\orange{(case 2.2)}}
    \end{aligned}
\end{equation*}

Noting that $N_{i(j)}(t_1, u_1)N_{i(j)}(t_2, u_2)$ jumps only when $t_1=t_2, u_1=u_2$, we again use the law of total expectations to simplify this expression:
\begin{equation*}
    \begin{aligned}
        &\EX[ h \bm{U}_i(v,\bm{\beta}(v))\bm{U}_i(v,\bm{\beta}(v))^{\text{T}}|R_i=r] \\
        &= \Bigg\{ \EX\bigg[ 
         r^{-2} 
         \sum_{j=1}^{r}\bigg\{\int_{0}^{1} \int_{0}^{\tau} h
        K_h^2(u_1-v) 
        \bigg[ \bm{Z}_i - \frac{s^{(1)}(t_1,\bm{\beta}(v))}{s^{(0)}(t_1,\bm{\beta}(v))}
    \bigg] 
    \bigg[ \bm{Z}_i - \frac{s^{(1)}(t_1,\bm{\beta}(v))}{s^{(0)}(t_1,\bm{\beta}(v))} \bigg]^\text{T} \\
    & \qquad \qquad\qquad\qquad \times f(t_1,u_1|T_{i(j)} \geq t_1, \bm{Z}_i, R_i=r) dt_1 du_1 \bigg\} \bigg| R_i =r \bigg] %\blue{\text{ (pt a)}}
    \\& \qquad \quad -  \EX\bigg[ 
        r^{-2} 
         \sum_{j=1}^{r}\bigg\{\int_{0}^{1} \int_{0}^{\tau} \int_{0}^{1} \int_{0}^{\tau} h
        K_h(u_1-v)  K_h(u_2-v) 
        \bigg[ \bm{Z}_i - \frac{s^{(1)}(t_1,\bm{\beta}(v))}{s^{(0)}(t_1,\bm{\beta}(v))}
    \bigg] 
    \bigg[ \bm{Z}_i - \frac{s^{(1)}(t_2,\bm{\beta}(v))}{s^{(0)}(t_2,\bm{\beta}(v))} \bigg]^\text{T} \\ &\qquad \qquad\qquad\qquad \times Y_{i(j)}(t_1) f(t_1,u_1|T_{i(j)} \geq t_1, \bm{Z}_i, R_i = r) Y_{i(j)}(t_2) \lambda(t_2,u_2|\bm{Z}_i) dt_1 du_1 dt_2 du_2 \bigg\}  \bigg| R_i = r \bigg] %\text{\blue{(pt b)}}
    \\& \qquad \quad -  \EX\bigg[ 
        r^{-2} 
         \sum_{j=1}^{r}\bigg\{\int_{0}^{1} \int_{0}^{\tau} \int_{0}^{1} \int_{0}^{\tau} h
        K_h(u_1-v)  K_h(u_2-v) 
        \bigg[ \bm{Z}_i - \frac{s^{(1)}(t_1,\bm{\beta}(v))}{s^{(0)}(t_1,\bm{\beta}(v))}
    \bigg] 
    \bigg[ \bm{Z}_i - \frac{s^{(1)}(t_2,\bm{\beta}(v))}{s^{(0)}(t_2,\bm{\beta}(v))} \bigg]^\text{T} \\ &\qquad \qquad\qquad\qquad \times Y_{i(j)}(t_2) f(t_2,u_2|T_{i(j)} \geq t_2, \bm{Z}_i, R_i = r) Y_{i(j)}(t_1) \lambda(t_1,u_1|\bm{Z}_i) dt_1 du_1 dt_2 du_2 \bigg\}  \bigg| R_i = r \bigg] %\blue{\text{ (pt c)}}
    \\ &\qquad \quad + \EX\bigg[
         r^{-2}  
         \sum_{j=1}^{r} \bigg\{ \int_{0}^{1} \int_{0}^{\tau} \int_{0}^{1} \int_{0}^{\tau} h
        K_h(u_1-v)  K_h(u_2-v) 
        \bigg[ \bm{Z}_i - \frac{s^{(1)}(t_1,\bm{\beta}(v))}{s^{(0)}(t_1,\bm{\beta}(v))} 
    \bigg] 
    \bigg[ \bm{Z}_i - \frac{s^{(1)}(t_2,\bm{\beta}(v))}{s^{(0)}(t_2,\bm{\beta}(v))} \bigg]^\text{T} \\ &\qquad \qquad\qquad\qquad \times Y_{i(j)}(t_1) \lambda(t_1,u_1|\bm{Z}_i) dt_1 du_1 Y_{i(j)}(t_2) \lambda(t_2,u_2|\bm{Z}_i) dt_2 du_2
        \bigg\} \bigg| R_i = r
        \bigg] %\text{\blue{(pt d)}}
        \Bigg\} \\
        & \quad + \Bigg\{ 
        \EX\bigg[
         r^{-2} \sum_{j\neq k}^{r} 
         \sum_{k=1}^{r}\bigg\{\int_{0}^{1} \int_{0}^{\tau} \int_{0}^{1} \int_{0}^{\tau} h
        K_h(u_1-v)  K_h(u_2-v) 
        \bigg[ \bm{Z}_i - \frac{s^{(1)}(t_1,\bm{\beta}(v))}{s^{(0)}(t_1,\bm{\beta}(v))}
    \bigg] 
    \bigg[ \bm{Z}_i - \frac{s^{(1)}(t_2,\bm{\beta}(v))}{s^{(0)}(t_2,\bm{\beta}(v))} \bigg]^\text{T} \\ &\qquad \qquad\qquad\qquad \times  Y_{i(j)}(t_1) Y_{i(k)}(t_2)f(t_1,u_1,t_2,u_2|T_{i(j)} \geq t_1, T_{i(k)} \geq t_2, \bm{Z}_i,R_i=r) dt_1 du_1 dt_2 du_2 \bigg\} \bigg| R_i =r\bigg] %\text{\blue{(pt a)}}
    \\& \qquad \quad - \EX\bigg[
         r^{-2} \sum_{j\neq k}^{r} 
         \sum_{k=1}^{r}\bigg\{\int_{0}^{1} \int_{0}^{\tau} \int_{0}^{1} \int_{0}^{\tau} h
        K_h(u_1-v)  K_h(u_2-v) 
        \bigg[ \bm{Z}_i - \frac{s^{(1)}(t_1,\bm{\beta}(v))}{s^{(0)}(t_1,\bm{\beta}(v))}
    \bigg] 
    \bigg[ \bm{Z}_i - \frac{s^{(1)}(t_2,\bm{\beta}(v))}{s^{(0)}(t_2,\bm{\beta}(v))} \bigg]^\text{T} \\ &\qquad \qquad\qquad\qquad \times Y_{i(j)}(t_1)f(t_1,u_1|T_{i(j)} \geq t_1, \bm{Z}_i, R_i=r)  Y_{i(k)}(t_2) \lambda(t_2,u_2|\bm{Z}_i) dt_1 du_1 dt_2 du_2 \bigg\} \bigg| R_i=r \bigg] %\text{\blue{(pt b)}}
    \\& \qquad \quad - \EX \bigg[
         r^{-2} \sum_{j\neq k}^{r} 
         \sum_{k=1}^{r}\bigg\{\int_{0}^{1} \int_{0}^{\tau} \int_{0}^{1} \int_{0}^{\tau} h
        K_h(u_1-v)  K_h(u_2-v) 
        \bigg[ \bm{Z}_i - \frac{s^{(1)}(t_1,\bm{\beta}(v))}{s^{(0)}(t_1,\bm{\beta}(v))}
    \bigg] 
    \bigg[ \bm{Z}_i - \frac{s^{(1)}(t_2,\bm{\beta}(v))}{s^{(0)}(t_2,\bm{\beta}(v))} \bigg]^\text{T}  \\ &\qquad \qquad\qquad\qquad \times Y_{i(k)}(t_2) f(t_2,u_2|T_{i(k)} \geq t_2, \bm{Z}_i, R_i=r) Y_{i(j)}(t_1) \lambda(t_1,u_1|\bm{Z}_i)dt_1 du_1 dt_2 du_2 \bigg\}  \bigg| R_i =r \bigg] %\text{\blue{(pt c)}}
    \\ & \qquad \quad + \EX\bigg[ \EX \bigg[
         r^{-2} \sum_{j\neq k}^{r} 
         \sum_{k=1}^{r} \bigg\{ \int_{0}^{1} \int_{0}^{\tau} \int_{0}^{1} \int_{0}^{\tau} h
        K_h(u_1-v)  K_h(u_2-v) 
        \bigg[ \bm{Z}_i - \frac{s^{(1)}(t_1,\bm{\beta}(v))}{s^{(0)}(t_1,\bm{\beta}(v))} 
    \bigg] 
    \bigg[ \bm{Z}_i - \frac{s^{(1)}(t_2,\bm{\beta}(v))}{s^{(0)}(t_2,\bm{\beta}(v))} \bigg]^\text{T} \\ &\qquad \qquad\qquad\qquad \times Y_{i(j)}(t_1) \lambda(t_1,u_1|\bm{Z}_i) Y_{i(k)}(t_2) \lambda(t_2,u_2|\bm{Z}_i) dt_1 du_1 dt_2 du_2
        \bigg\} \bigg| R_i =r
        \bigg] %\text{\blue{(pt d)}}
        \Bigg\}
    \end{aligned}
\end{equation*}

Then, using u-substitutions with $u_1' = \frac{u_1-v}{h}, u_2' = \frac{u_2-v}{h}$ and taking the limit as $h \rightarrow 0$, we get:

\begin{equation*}
    \begin{aligned}
    &\EX[ h \bm{U}_i(v,\bm{\beta}(v))\bm{U}_i(v,\bm{\beta}(v))^{\text{T}}|R_i=r] \\  & \qquad \rightarrow \nu_0 \EX\bigg[
         r^{-2}  
         \sum_{j=1}^{r} \bigg\{\int_{0}^{\tau} 
        \bigg[ \bm{Z}_i - \frac{s^{(1)}(t_1,\bm{\beta}(v))}{s^{(0)}(t_1,\bm{\beta}(v))}
    \bigg] 
    \bigg[ \bm{Z}_i - \frac{s^{(1)}(t_1,\bm{\beta}(v))}{s^{(0)}(t_1,\bm{\beta}(v))} \bigg]^\text{T} f(t_1,v|T_{i(j)} \geq t_1, \bm{Z}_i, R_i = r) dt_1  \bigg\} \bigg| R_i = r \bigg] \\
    & \qquad \quad (\equiv \bm{\Lambda}_r(v)).
    \end{aligned}
\end{equation*}

Putting everything together, we have 
\begin{equation*}
    \begin{aligned}
        \EX[ h \bm{U}_i(v,\bm{\beta}(v))\bm{U}_i(v,\bm{\beta}(v))^{\text{T}}]
        &= \sum_{r=1}^{\infty}\Pr(R_i=r)\bm{\Lambda}_r(v) \equiv \bm{\Lambda}(v).
    \end{aligned}
\end{equation*}
Then $Var(n^{-1/2}h^{1/2}\bm{U}(v,\bm{\beta}(v)) = hVar(\bm{U}_i(v,\bm{\beta}(v))) = \bm{\Lambda}(v)$ %\pink{(??? Why still an $h$ here?)} 
and $\sqrt{nh}(\bm{U}(v,\bm{\beta}(v)) \xrightarrow{D} N(0, \bm{\Lambda}(v))$.

%\textbf{\textit{Step 4: Sandwich}}
%\blue{
Note that 
\[
    \begin{aligned}
        \bm{U}(v,\hat{\bm{\beta}}(v))-\bm{U}(v,\bm{\beta}(v)) &=  
        \int_{0}^{1} \nabla_{ \bm{\beta}} \bm{U}(v, (\bm{\beta}(v) +u(\hat{\bm{\beta}}(v)-\bm{\beta}(v)))) du \times(\hat{\bm{\beta}}(v)-\bm{\beta}(v)) \\
        &= \int_{0}^{1} \nabla^2_{ \bm{\beta} \bm{\beta}} l(v, (\bm{\beta}(v) +u(\hat{\bm{\beta}}(v)-\bm{\beta}(v)))) du \times(\hat{\bm{\beta}}(v)-\bm{\beta}(v)). 
    \end{aligned}
\]
Then for $u \in (0,1)$, we have
\[
    \begin{aligned}
        \hat{\bm{\beta}}(v)-\bm{\beta}(v) = - \bigg( \int_{0}^{1} \nabla^2_{ \bm{\beta} \bm{\beta}} l(v, (\bm{\beta}(v) +u(\hat{\bm{\beta}}(v)-\bm{\beta}(v)))) du \bigg)^{-1}(\bm{U}(v,\bm{\beta}(v)))
    \end{aligned}
\]

By \textbf{Lemma 7.8} and \textbf{Proposition 7.9} from \cite{yi_li_lecture_nodate}, the regularity conditions, and the uniform consistency of $\hat{\bm{\beta}}(v)$ for $v \in [a,b] \subset (0, 1)$, we can establish that $ \int_{0}^{1} 
\frac{1}{n}\nabla^2_{ \bm{\beta} \bm{\beta}} l(v, (\bm{\beta}(v) +u(\hat{\bm{\beta}}(v)-\bm{\beta}(v)))) du\xrightarrow{p} \bm{\Sigma}(v)$ (as defined in the regularity conditions) uniformly in $v \in [a,b]$ for $0<\delta<1/2$.

\sloppy Putting everything together, since $- \bigg( \int_{0}^{1}  \frac{1}{n} \nabla^2_{ \bm{\beta} \bm{\beta}} l(v, (\bm{\beta}(v) +u(\hat{\bm{\beta}}(v)-\bm{\beta}(v)))) du \bigg)^{-1} \xrightarrow{p} -\bm{\Sigma}(v)^{-1}$ uniformly in $v \in [a,b]$ for $0<\delta<1/2$ by the Continuous Mapping theorem, and $(n^{1/2}h^{1/2} \bm{U}(v,\bm{\beta}(v))) \xrightarrow{D} N(0,\bm{\Lambda}(v))$, then, by Slutsky's theorem, we have:
\[
    \begin{aligned}
        \sqrt{nh}(\hat{\bm{\beta}}(v)-\bm{\beta}(v)) \xrightarrow{D} N(0,\bm{\Sigma}(v)^{-1} \bm{\Lambda}(v) \bm{\Sigma}(v)^{-1} ) \text{ as $nh^3 \rightarrow \infty$ and $nh^5 \rightarrow 0$}. 
    \end{aligned}
\]

$\qedsymbol$

\section{Full Modeling Results: Mark-Specific Hazards of Opioid Refill}
\begin{landscape}
\begingroup
\renewcommand\arraystretch{0.6}
\begin{longtable}{lcccccccccccc}
\caption{Mark-specific recurrent events proportional hazards model results for post-surgical opioid prescription refills at dosages between 50 to 450 OMEs, compared to non-mark-specific recurrent events proportional hazards model results} \label{tab:opioid_modelres} \\     
\hline
 & \multicolumn{10}{c}{\textbf{Mark-Specific Model$^1$}} & \multicolumn{2}{c}{\textbf{Non}} \\ \cline{2-11} 
 &  \multicolumn{10}{c}{\textbf{Dosage}} & \multicolumn{2}{c}{\textbf{Mark-Specific}}\\ \cline{2-11}
 &  \multicolumn{2}{c}{\textbf{50 OMEs}} & \multicolumn{2}{c}{\textbf{100 OMEs}} & \multicolumn{2}{c}{\textbf{150 OMEs}} & \multicolumn{2}{c}{\textbf{300 OMEs}} & \multicolumn{2}{c}{\textbf{450 OMEs}}  & \multicolumn{2}{c}{\textbf{Model$^1$}}\\ \cline{2-11}
 \textbf{Covariates} & \textbf{HR} & \textbf{95\% CI}  & \textbf{HR} & \textbf{95\% CI} & \textbf{HR} & \textbf{95\% CI} & \textbf{HR} & \textbf{95\% CI} & \textbf{HR} & \textbf{95\% CI} & \textbf{HR} & \textbf{95\% CI} \\ \hline

\endfirsthead

\multicolumn{13}{c}%
{{\itshape S\tablename\ \thetable{} -- continued from previous page}} \\
\hline
 & \multicolumn{10}{c}{\textbf{Mark-Specific Model$^1$}} & \multicolumn{2}{c}{\textbf{Non}} \\ \cline{2-11} 
 &  \multicolumn{10}{c}{\textbf{Dosage}} & \multicolumn{2}{c}{\textbf{Mark-Specific}}\\ \cline{2-11}
 &  \multicolumn{2}{c}{\textbf{50 OMEs}} & \multicolumn{2}{c}{\textbf{100 OMEs}} & \multicolumn{2}{c}{\textbf{150 OMEs}} & \multicolumn{2}{c}{\textbf{300 OMEs}} & \multicolumn{2}{c}{\textbf{450 OMEs}}  & \multicolumn{2}{c}{\textbf{Model$^1$}}\\ \cline{2-11}
 \textbf{Covariates} & \textbf{HR} & \textbf{95\% CI}  & \textbf{HR} & \textbf{95\% CI} & \textbf{HR} & \textbf{95\% CI} & \textbf{HR} & \textbf{95\% CI} & \textbf{HR} & \textbf{95\% CI} & \textbf{HR} & \textbf{95\% CI} \\ \hline
\endhead

\hline \multicolumn{13}{r}{{Continued on next page}} \\ \hline
\endfoot

\endlastfoot

    Sex (ref=Female) & & & & & & & & & & & &\\
    \hspace{1em} Male & 0.985 & (0.886, 1.096) & \textbf{0.862} & \textbf{(0.772, 0.963)} & 0.966 & (0.846, 1.102) & 1.057 & (0.932, 1.199) & 1.041 & (0.929, 1.167) & 0.973 & (0.916, 1.034) \\
    
    Race (ref=White)  & & & & & & & & & & & &\\
    \hspace{1em} Black & 1.161 & (0.998, 1.350) & \textbf{1.199} & \textbf{(1.045, 1.376)} & 1.015 & (0.855, 1.204) & \textbf{0.792} & \textbf{(0.654, 0.960)} & 0.907 & (0.767, 1.073) & \textbf{1.109} & \textbf{(1.023, 1.202)} \\
    \hspace{1em} Asian & 0.917 & (0.561, 1.499) & 0.600 & (0.325, 1.105) & 1.230 & (0.733, 2.064) & 0.908 & (0.485, 1.700) & 0.718 & (0.369, 1.398) & 0.909 & (0.667, 1.239) \\
    \hspace{1em} Other & 1.120 & (0.572, 2.192) & 1.027 & (0.538, 1.962) & 0.636 & (0.237, 1.705) & 0.844 & (0.382, 1.864) & 0.896 & (0.435, 1.849) & 1.002 & (0.691, 1.452) \\
    
    Age (ref=55-$<$65)  & & & & & & & & & & & &\\
    \hspace{1em} 18-$<$25 & 1.241 & (0.980, 1.571) & \textbf{1.428} & \textbf{(1.118, 1.824)} & 0.988 & (0.684, 1.428) & \textbf{0.481} & \textbf{(0.300, 0.772)} & \textbf{0.388} & \textbf{(0.233, 0.648)} & \textbf{1.177} & \textbf{(1.001, 1.384)} \\
    \hspace{1em} 25-$<$35 & \textbf{1.269} & \textbf{(1.068, 1.509)} & 1.147 & (0.959, 1.372) & 0.976 & (0.772, 1.234) & \textbf{0.565} & \textbf{(0.432, 0.740)} & \textbf{0.513} & \textbf{(0.390, 0.675)} & 1.061 & (0.951, 1.184) \\
    \hspace{1em} 35-$<$45 & 1.155 & (0.983, 1.357) & 1.050 & (0.898, 1.227) & 1.180 & (0.980, 1.421) & 0.954 & (0.787, 1.155) & 0.837 & (0.688, 1.017) & 1.086 & (0.990, 1.192) \\
    \hspace{1em} 45-$<$55 & 0.909 & (0.783, 1.055) & 1.023 & (0.891, 1.174) & 1.169 & (0.993, 1.376) & 0.866 & (0.732, 1.023) & 0.885 & (0.754, 1.040) & 0.988 & (0.911, 1.072) \\
    \hspace{1em} 65-$<$75 & \textbf{0.824} & \textbf{(0.714, 0.951)} & 0.916 & (0.799, 1.050) & 1.046 & (0.891, 1.226) & 0.995 & (0.858, 1.154) & 1.062 & (0.925, 1.219) & \textbf{0.915} & \textbf{(0.848, 0.987)} \\
    \hspace{1em} 75+ & \textbf{0.807} & \textbf{(0.686, 0.949)} & 0.795 & \textbf{(0.675, 0.936)} & 0.834 & (0.689, 1.010) & 0.860 & (0.720, 1.026) & 1.145 & (0.981, 1.336) & \textbf{0.826} & \textbf{(0.757, 0.902)} \\

    BMI (ref=25-$<$30)   & & & & & & & & & & & &\\
    \hspace{1em} $\leq$25 & 1.076 & (0.950, 1.217) & 0.899 & (0.795, 1.017) & 1.009 & (0.873, 1.166) & \textbf{0.853} & \textbf{(0.735, 0.990)} & 1.021 & (0.892, 1.168) & 0.991 & (0.923, 1.064) \\
    \hspace{1em} 30 - $<$35 & 1.080 & (0.954, 1.221) & 0.934 & (0.827, 1.054) & 1.017 & (0.880, 1.175) & 1.059 & (0.918, 1.221) & 1.090 & (0.952, 1.249) & 1.037 & (0.967, 1.113) \\
    \hspace{1em} 35 - $<$40 & 0.994 & (0.849, 1.163) & 1.098 & (0.953, 1.266) & 1.128 & (0.949, 1.342) & 1.177 & (0.988, 1.403) & \textbf{1.186} & \textbf{(1.002, 1.404)} & 1.088 & (0.999, 1.185) \\
    \hspace{1em} 40 - $<$45 & 1.164 & (0.956, 1.416) & 0.991 & (0.817, 1.203) & 1.046 & (0.828, 1.321) & \textbf{1.268} & \textbf{(1.008, 1.595)} & \textbf{1.369} & \textbf{(1.099, 1.705)} & 1.087 & (0.969, 1.220) \\
    \hspace{1em} 45 - $<$50 & 0.982 & (0.717, 1.346) & 1.252 & (0.970, 1.617) & 1.162 & (0.835, 1.618) & 1.205 & (0.847, 1.715) & \textbf{1.785} & \textbf{(1.334, 2.387)} & \textbf{1.240} & \textbf{(1.060, 1.450)} \\
    \hspace{1em} 50+ & 0.682 & (0.456, 1.019) & 1.031 & (0.759, 1.399) & \textbf{1.488} & \textbf{(1.072, 2.066)} & \textbf{1.906} & \textbf{(1.377, 2.640)} & \textbf{2.096} & \textbf{(1.540, 2.853)} & 1.131 & (0.940, 1.360) \\

Ascites Status (ref=No)  & & & & & & & & & & & &\\
\hspace{1em} Yes & 1.082 & (0.606, 1.933) & 1.106 & (0.655, 1.867) & 0.939 & (0.508, 1.736) & 1.099 & (0.638, 1.893) & \textbf{1.645} & \textbf{(1.116, 2.424)} & \textbf{1.281} & \textbf{(1.044, 1.572)} \\

Cancer Status (ref=No)  & & & & & & & & & & & &\\
\hspace{1em} Yes  & 0.835 & (0.688, 1.014) & 0.931 & (0.792, 1.094) & 0.992 & (0.828, 1.190) & \textbf{1.467} & \textbf{(1.243, 1.733)} & \textbf{1.739} & \textbf{(1.502, 2.015)} & 1.071 & (0.982, 1.168) \\

Chronic Condition Status (ref=No)  & & & & & & & & & & & &\\
\hspace{1em} Yes & 1.049 & (0.534, 2.060) & 0.884 & (0.434, 1.801) & 0.751 & (0.325, 1.734) & 1.232 & (0.706, 2.150) & 1.102 & (0.701, 1.732) & 1.099 & (0.855, 1.413) \\

COPD (ref=No) & & & & & & & & & & & &\\
\hspace{1em} Yes  & \textbf{0.697} & \textbf{(0.501, 0.968)} & \textbf{0.701} & \textbf{(0.514, 0.955)} & 1.035 & (0.767, 1.395) & 1.138 & (0.861, 1.505) & \textbf{1.305} & \textbf{(1.031, 1.650)} & 0.961 & (0.837, 1.104) \\

Congestive Heart Failure (ref=No) & & & & & & & & & & & &\\
\hspace{1em} Yes  & 0.870 & (0.692, 1.095) & \textbf{0.771} & \textbf{(0.610, 0.975)} & 0.835 & (0.647, 1.077) & 1.021 & (0.825, 1.264) & \textbf{1.317} & \textbf{(1.112, 1.560)} & 0.984 & (0.881, 1.100) \\ \\

Diabetes (ref=No)  & & & & & & & & & & & &\\
\hspace{1em} Type I Diabetes & 1.070 & (0.858, 1.335) & \textbf{0.705} & \textbf{(0.552, 0.900)} & 1.051 & (0.827, 1.335) & 0.889 & (0.700, 1.129) & 0.955 & (0.783, 1.165) & 0.934 & (0.831, 1.049) \\
\hspace{1em} Type II Diabetes & 1.068 & (0.910, 1.253) & 1.054 & (0.906, 1.226) & 0.991 & (0.828, 1.187) & 0.954 & (0.803, 1.132) & 0.863 & (0.736, 1.012) & 1.008 & (0.925, 1.098) \\

Dialysis (ref=No)  & & & & & & & & & & & &\\
\hspace{1em} Yes & 0.724 & (0.366, 1.434) & 1.168 & (0.643, 2.122) & 0.632 & (0.283, 1.415) & 0.795 & (0.414, 1.525) & 0.860 & (0.513, 1.440) & 0.867 & (0.652, 1.154) \\

Functional Status (ref=Independent)  & & & & & & & & & & & &\\
\hspace{1em} Not Independent & \textbf{0.463} & \textbf{(0.299, 0.717)} & \textbf{0.326} & \textbf{(0.192, 0.554)} & 0.865 & (0.594, 1.260) & 0.852 & (0.594, 1.222) & 1.209 & (0.939, 1.557) & \textbf{0.683} & \textbf{(0.566, 0.823)} \\ 
\hspace{1em} Unknown & 0.836 & (0.297, 2.350) & 0.382 & (0.092, 1.583) & 0.504 & (0.108, 2.344) & 1.495 & (0.641, 3.485) & 0.810 & (0.323, 2.036) & 0.757 & (0.424, 1.351) \\

Hypertension Status (ref=No)  & & & & & & & & & & & &\\
\hspace{1em} Yes & 1.032 & (0.923, 1.153) & 0.982 & (0.884, 1.092) & 1.029 & (0.909, 1.163) & \textbf{1.146} & \textbf{(1.014, 1.294)} & \textbf{1.261} & \textbf{(1.126, 1.413)} & 1.023 & (0.963, 1.086) \\

Preoperative Sepsis Status (ref=No)   & & & & & & & & & & & &\\
\hspace{1em} Yes & 0.913 & (0.748, 1.115) & 1.175 & (0.956, 1.445) & 1.031 & (0.791, 1.345) & \textbf{1.408} & \textbf{(1.103, 1.797)} & \textbf{1.262} & \textbf{(1.016, 1.567)} & 1.012 & (0.899, 1.140) \\
\hspace{1em} Yes (Severe) & 0.860 & (0.628, 1.177) & 0.934 & (0.666, 1.309) & 0.848 & (0.579, 1.242) & 1.065 & (0.754, 1.505) & 0.836 & (0.617, 1.131) & 0.886 & (0.757, 1.037) \\

Sleep Apnea Status (ref=No)  & & & & & & & & & & & &\\
\hspace{1em} Yes & 0.983 & (0.869, 1.113) & \textbf{1.130} & \textbf{(1.006, 1.269)} & 0.970 & (0.844, 1.115) & 0.884 & (0.774, 1.010) & 0.980 & (0.870, 1.105) & 1.002 & (0.938, 1.070) \\

Smoking Status (ref=No)  & & & & & & & & & & & &\\
\hspace{1em} Yes & 1.037 & (0.921, 1.168) & 1.099 & (0.981, 1.232) & 1.142 & (0.999, 1.307) & \textbf{1.510} & \textbf{(1.326, 1.720)} & \textbf{1.669} & \textbf{(1.479, 1.883)} & \textbf{1.174} & \textbf{(1.101, 1.252)} \\

Ventilator Status (ref=No)  & & & & & & & & & & & &\\
\hspace{1em} Yes  & 0.935 & (0.396, 2.206) & 0.472 & (0.138, 1.615) & 0.760 & (0.298, 1.936) & 1.012 & (0.479, 2.139) & 1.018 & (0.577, 1.797) & 0.761 & (0.521, 1.113) \\

Surgery Type (ref=Minor Hernia)  & & & & & & & & & & & &\\
\hspace{1em} Abdominal Hernia & \textbf{0.708} & \textbf{(0.516, 0.970)} & 0.878 & (0.660, 1.168) & 1.168 & (0.863, 1.581) & 1.081 & (0.816, 1.433) & 1.012 & (0.774, 1.325) & 0.928 & (0.799, 1.079) \\
\hspace{1em} Carotid Endarterectomy & 0.714 & (0.417, 1.222) & 0.694 & (0.405, 1.186) & \textbf{0.432} & \textbf{(0.211, 0.885)} & \textbf{0.472} & \textbf{(0.251, 0.887)} & \textbf{0.307} & \textbf{(0.153, 0.616)} & \textbf{0.591} & \textbf{(0.439, 0.796)} \\[-3pt]
\hspace{1em} \multirow{2}{0.3\textwidth}{\vspace{-1em}Creation, Re-siting, or Closure of Ileostomy or Colostomy} & & & & & & & & & & & &\\
& 0.777 & (0.475, 1.270) & 0.806 & (0.505, 1.285) & 0.753 & (0.441, 1.286) & 0.972 & (0.626, 1.510) & 1.244 & (0.867, 1.783) & 0.835 & (0.678, 1.028) \\ 
\hspace{1em} \multirow{2}{0.3\textwidth}{\vspace{-1em}Laparoscopic Anti-Reflux and Hiatal Hernia Surgery}  & & & & & & & & & & & &\\ 
& 1.047 & (0.706, 1.554) & 1.194 & (0.835, 1.706) & 0.935 & (0.582, 1.500) & 1.013 & (0.611, 1.680) & 1.061 & (0.667, 1.688) & 1.006 & (0.791, 1.278) \\ \\[-7pt]
\hspace{1em} Laparoscopic Appendectomy & 1.191 & (0.953, 1.489) & 0.980 & (0.773, 1.242) & 0.807 & (0.599, 1.088) & 0.940 & (0.695, 1.272) & \textbf{0.698} & \textbf{(0.526, 0.928)} & 1.074 & (0.937, 1.231) \\
\hspace{1em} Laparoscopic Cholecystectomy & 1.078 & (0.898, 1.296) & 0.872 & (0.720, 1.055) & 0.835 & (0.661, 1.054) & 1.139 & (0.912, 1.421) & 0.985 & (0.806, 1.204) & 1.036 & (0.933, 1.150) \\
\hspace{1em} Laparoscopic Colectomy & 0.991 & (0.763, 1.286) & 1.072 & (0.837, 1.375) & 0.909 & (0.671, 1.231) & 1.231 & (0.922, 1.644) & 0.964 & (0.735, 1.265) & 0.953 & (0.829, 1.095) \\
\hspace{1em} Laparoscopic Hysterectomy & \textbf{0.677} & \textbf{(0.513, 0.893)} & \textbf{1.626} & \textbf{(1.300, 2.034)} & \textbf{1.706} & \textbf{(1.303, 2.233)} & 1.225 & (0.909, 1.650) & 0.992 & (0.743, 1.325) & \textbf{1.386} & \textbf{(1.203, 1.598)} \\
\hspace{1em} Open Appendectomy & \textbf{1.596} & \textbf{(1.041, 2.447)} & 0.712 & (0.397, 1.277) & 0.680 & (0.341, 1.358) & 0.533 & (0.251, 1.132) & 0.886 & (0.527, 1.492) & 0.975 & (0.725, 1.311) \\
\hspace{1em} Open Cholecystectomy & 1.277 & (0.797, 2.047) & 0.997 & (0.591, 1.680) & 1.419 & (0.853, 2.358) & 0.557 & (0.272, 1.141) & 0.536 & (0.283, 1.016) & 1.028 & (0.802, 1.317) \\
\hspace{1em} Open Colectomy & 0.833 & (0.636, 1.092) & 1.018 & (0.788, 1.316) & 1.152 & (0.868, 1.529) & 0.944 & (0.720, 1.237) & 1.088 & (0.860, 1.377) & 0.951 & (0.836, 1.082) \\ \\
\hspace{1em} \multirow{2}{0.3\textwidth}{\vspace{-1em}Open Small Bowel Resection of Enterolysis}  & & & & & & & & & & & &\\ [-3pt]
& \textbf{0.529} & \textbf{(0.359, 0.778)} & 0.710 & (0.496, 1.018) & \textbf{0.646} & \textbf{(0.427, 0.977)} & 0.713 & (0.490, 1.305) & \textbf{0.716} & \textbf{(0.521, 0.985)} & \textbf{0.638} & \textbf{(0.531, 0.766)} \\ \\[-3pt]
\hspace{1em} Thyroidectomy & 1.234 & (0.899, 1.692) & 0.799 & (0.564, 1.132) & 0.765 & (0.443, 1.029) & 0.700 & (0.473, 1.034) & \textbf{0.650} & \textbf{(0.442, 0.957)} & 0.849 & (0.696, 1.036) \\
\hspace{1em} Total Abdominal Hysterectomy & \textbf{0.612} & \textbf{(0.409, 0.916)} & \textbf{2.202} & \textbf{(1.538, 2.659)} & \textbf{1.915} & \textbf{(1.387, 2.644)} & 1.716 & (1.23, 2.394) & \textbf{1.501} & \textbf{(1.074, 2.096)} & \textbf{1.385} & \textbf{(1.175, 1.631)} \\
\hspace{1em} Vaginal Hysterectomy & 0.973 & (0.734, 1.292) & 1.291 & (0.997, 1.669) & \textbf{1.776} & \textbf{(1.327, 2.378)} & 1.344 & (0.962, 1.878) & 1.104 & (0.788, 1.547) & \textbf{1.345} & \textbf{(1.138, 1.589)} \\
\hspace{1em} Other & \textbf{0.722} & \textbf{(0.551, 0.946)} & 0.791 & (0.607, 1.030) & 0.867 & (0.648, 1.160) & 1.064 & (0.816, 1.389) & 1.159 & (0.923, 1.456) & \textbf{0.866} & \textbf{(0.760, 0.987)} \\

Surgical Priority (ref=Elective)  & & & & & & & & & & & &\\
\hspace{1em} Urgent/Emergent & 1.088 & (0.940, 1.258) & 0.923 & (0.798, 1.068) & 0.904 & (0.764, 1.070) & 0.854 & (0.725, 1.007) & 1.154 & (1.000, 1.333) & 0.956 & (0.885, 1.032) \\
Surgery Location (ref=Outpatient)  & & & & & & & & & & & &\\
\hspace{1em} Inpatient & \textbf{3.903} & \textbf{(3.372, 4.517)} & \textbf{3.962} & \textbf{(3.437, 4.568)} & \textbf{4.550} & \textbf{(3.833, 5.403)} & \textbf{3.465} & \textbf{(2.926, 4.103)} & \textbf{2.903} & \textbf{(2.475, 3.404)} & \textbf{4.148} & \textbf{(3.823, 4.501)} \\[-7pt]
\multirow{2}{0.4\textwidth}{\vspace{-1em}Surgical Approach (ref=Minimally Invasive)}  & & & & & & & & & & & &\\ \\
\hspace{1em} Open & 0.931 & (0.768, 1.127) & 0.967 & (0.797, 1.173) & 1.068 & (0.854, 1.337) & \textbf{1.361} & \textbf{(1.107, 1.673)} & 1.156 & (0.964, 1.387) & 1.039 & (0.938, 1.151) \\
\multirow{2}{0.35\textwidth}{\vspace{-1em}No. of Preoperative Prescriptions - Benzo} & & & & & & & & & & & & \\
& 1.004 & (0.935, 1.078) & 1.060 & (1.000, 1.124) & 1.046 & (0.979, 1.119) & 0.963 & (0.891, 1.041) & 1.015 & (0.951, 1.083) & 1.024 & (0.988, 1.061) \\
No. of Preoperative Fills - Benzo  & 1.000 & (0.999, 1.001) & 1.000 & (0.999, 1.001) & 1.000 & (0.999, 1.001) & 1.001 & (1.000, 1.002) & 1.000 & (0.999, 1.001) & 1.000 & (0.999, 1.000) \\
No. of Prescribers by Specialty - Benzo  & 0.895 & (0.616, 1.299) & 0.908 & (0.643, 1.281) & 1.116 & (0.772, 1.615) & 1.028 & (0.682, 1.550) & 1.262 & (0.889, 1.790) & 1.005 & (0.834, 1.211) \\
No. of Pharmacies - Benzo  & 0.910 & (0.649, 1.277) & \textbf{0.612} & \textbf{(0.428, 0.874)} & 0.712 & (0.484, 1.047) & 0.854 & (0.578, 1.262) & 0.792 & (0.546, 1.151) & \textbf{0.758} & \textbf{(0.625, 0.919)} \\
No. of Drug Products - Benzo  & 1.226 & (0.793, 1.895) & 1.425 & (0.956, 2.125) & 1.088 & (0.690, 1.716) & 1.025 & (0.619, 1.697) & 0.840 & (0.530, 1.333) & 1.155 & (0.921, 1.449) \\
Average Days' Supply - Benzo  & 1.003 & (0.996, 1.009) & 1.003 & (0.996, 1.010) & 1.003 & (0.996, 1.011) & 1.000 & (0.993, 1.008) & 1.000 & (0.992, 1.007) & 1.002 & (0.998, 1.006) \\
\multirow{2}{0.35\textwidth}{\vspace{-1em}No. of Preoperative Prescriptions - Gaba}  & & & & & & & & & & & &\\
& 0.914 & (0.714, 1.171) & 0.927 & (0.728, 1.179) & 1.089 & (0.882, 1.344) & 0.902 & (0.629, 1.294) & 0.979 & (0.784, 1.223) & 0.966 & (0.836, 1.117) \\
No. of Preoperative Fills - Gaba & 1.001 & (1.000, 1.002) & 1.000 & (0.998, 1.002) & 0.999 & (0.996, 1.001) & 0.999 & (0.996, 1.002) & 1.000 & (0.998, 1.002) & 1.000 & (0.999, 1.001) \\
No. of Prescribers by Specialty - Gaba & 0.976 & (0.389, 2.451) & 1.837 & (0.868, 3.886) & 1.042 & (0.488, 2.225) & 1.111 & (0.310, 3.982) & 0.495 & (0.145, 1.689) & 1.136 & (0.759, 1.702) \\
No. of Pharmacies - Gaba & 0.830 & (0.305, 2.259) & 0.441 & (0.119, 1.630) & 1.798 & (0.897, 3.605) & 0.698 & (0.147, 3.323) & 0.765 & (0.238, 2.459) & 0.920 & (0.576, 1.468) \\
No. of Drug Products - Gaba & 1.011 & (0.258, 3.967) & 2.082 & (0.538, 8.062) & 0.394 & (0.128, 1.216) & 1.892 & (0.311, 11.512) & 3.734 & (0.720, 19.350) & 1.002 & (0.519, 1.934) \\
Average Days' Supply - Gaba & 1.003 & (0.991, 1.016) & 0.994 & (0.981, 1.007) & 1.005 & (0.990, 1.021) & 0.996 & (0.979, 1.014) & 0.998 & (0.977, 1.004) & 0.998 & (0.989, 1.006) \\
        \hline \\[-10pt]
\multicolumn{13}{l}{$^{\bm{1}}$Statistically significant results at the $\alpha$=0.05 level are bolded.}
    \end{longtable}
\endgroup
\end{landscape}

\bibliography{references_supp}